\documentclass[a4paper,12pt]{article}
\usepackage{latexsym}
\usepackage{amssymb}
\usepackage{amsmath}
\usepackage{cite}
\usepackage{color}

\usepackage{microtype}
\usepackage{slashed}

\numberwithin{equation}{section}
\allowdisplaybreaks

\makeatletter \@addtoreset{equation}{section} \makeatother

\usepackage{cancel}
\def\ii{{\rm i}}
\usepackage{ulem}
\usepackage{ytableau}

\ytableausetup{centertableaux, mathmode, smalltableaux}

\definecolor{blue-violet}{rgb}{0.54, 0.17, 0.89}
\definecolor{PineGreen}{cmyk}{0.92, 0, 0.59, 0.25}
\definecolor{YellowOrange}{cmyk}{0, 0.42, 1, 0}

\newcommand{\diff}{\mathrm{d}}

\begin{document}
\numberwithin{equation}{section}


\begin{center}
{\bf\LARGE On the geometric approach to the boundary problem in supergravity} \\
\vskip 2 cm
{\bf \large Laura Andrianopoli$^{1,2,3}$ and Lucrezia Ravera$^{1,2}$}
\vskip 8mm
 \end{center}
\noindent {\small $^{1}$ \it DISAT, Politecnico di Torino, Corso Duca degli Abruzzi 24, 10129 Torino, Italy. \\
    $^{2}$ \it INFN, Sezione di Torino, Via P. Giuria 1, 10125 Torino, Italy. \\
    $^{1}$ \it Arnold-Regge Center, Via P. Giuria 1, 10125 Torino, Italy.
}

\vskip 2 cm
\begin{center}
{\small {\bf Abstract}}
\end{center}

We review the geometric superspace approach to the boundary problem in supergravity, retracing the geometric construction of  four-dimensional supergravity Lagrangians in the presence of a non-trivial boundary of spacetime. We first focus on  pure $\mathcal{N}=1$ and $\mathcal{N}=2$ theories with negative cosmological constant. Here, the supersymmetry invariance of the action requires the addition of topological (boundary) contributions which generalize at the supersymmetric level the Euler-Gauss-Bonnet term. Moreover, one finds that the boundary values of the super field-strengths are dynamically fixed to constant values, corresponding to the vanishing of the $\rm{OSp}(\mathcal{N}|4)$-covariant supercurvatures at the boundary.
We then consider the case of vanishing cosmological constant where, in the presence of a non-trivial boundary, the inclusion of boundary terms involving additional fields, which behave as auxiliary fields for the bulk theory, allows to restore supersymmetry. In all the cases listed above, the full, supersymmetric Lagrangian can be recast in a MacDowell-Mansouri(-like) form. We then report on the application of the results to specific problems regarding cases where the boundary is located asymptotically,  relevant for a holographic analysis. 

\vspace{0.5cm}


\vfill
\noindent {\small{\it
    E-mail:  \\
{\tt laura.andrianopoli@polito.it}; \\
{\tt lucrezia.ravera@polito.it}}}
   \eject

\tableofcontents

\noindent\hrulefill

\section{Preamble}

As is well known, Einstein's theory of general relativity describes gravitation in terms of fluctuations of the geometry of spacetime. In particular, the gravitational field is expressed by the metric of spacetime, and Lagrangians describing  gravitational systems have to be invariant under general coordinate transformations, associated with spacetime diffeomorphisms.
This fundamental property is shared by all  theories that include gravitation in their formulation, and then in particular by supergravity theories, where spacetime diffeomorphisms are generated when acting twice with local supersymmetry transformations. 

The presence of a spacetime defect, and in particular of a boundary, is problematic because it introduces a length scale in the theory and  breaks  diffeomorphisms invariance. To recover such invariance of the action it is necessary to modify the theory by adding appropriate boundary contributions to the Lagrangian.
This issue, both for gravity and supergravity Lagrangians, has been analyzed in different contexts from the early seventies on, after the pioneering works \cite{York:1972sj, Gibbons:1976ue}, in first attempts to study the quantization of gravity within a path integral approach.

The boundary problem in gravity was carefully analyzed in the geometric Cartan approach in \cite{Aros:1999id,Aros:1999kt,Mora:2004kb,Olea:2005gb,Jatkar:2014npa,Jatkar:2015ffa}. In this framework, it was shown that the diffeomorphisms invariance of the bulk Einstein Lagrangian plus cosmological constant, which is broken in the presence of a boundary, can be restored by adding the topological Euler-Gauss-Bonnet (EGB) term
\begin{equation}\label{EGBterm}
    \mathcal{L}_{\text{EGB}} = \mathcal{R}^{ab} \wedge \mathcal{R}^{cd} \epsilon_{abcd} = \diff \left( \omega^{ab} \wedge \mathcal{R}^{cd} + {\omega^a}_{{f}} \wedge \omega^{{f}b} \wedge \omega^{cd} \right) \epsilon_{abcd}
\end{equation}
to the bulk Lagrangian of the theory. This leads to a background-independent definition of Noether charges, without the need of explicitly imposing Dirichlet boundary conditions on the fields (cf. also \cite{Belyaev:2005rt,Belyaev:2007bg,Belyaev:2008ex,Grumiller:2009dx,Belyaev:2010as}). 

\vskip 5mm
In the supergravity context, spacetime geometries including defects and boundaries are quite ubiquitous, as supergravity  gives an effective field theory description at low energy (with respect to the Planck scale) of a microscopic quantum gravity theory described in terms of superstrings. Indeed, a non-perturbative formulation of superstring theory requires the inclusion of dynamical spacetime defects, the D-branes, which are BPS objects satisfying a no-force condition that allows them to back-react non-trivially on the target space geometry of the low-energy theory, as spacetime defects (black $p$-branes) and non-trivial boundaries.
This mechanism attracted further attention in the last 25 years, since it is at the heart of the formulation of the well-celebrated AdS$_{d+1}$/CFT$_d$ holographic duality  \cite{Maldacena:1997re}. Indeed,
the analysis of the black $p$-brane effective geometry
emerging from the superposition of a large number of   D3-branes in the microscopic theory, in the special limit considered in \cite{Maldacena:1997re} led  to  establish the \textit{holographic correspondence} \cite{Maldacena:1997re,Gubser:1998bc,Witten:1998qj,Aharony:1999ti}. The latter is a duality between supergravity on the near-horizon geometry of a black $p$-brane, that is on   asymptotically AdS$_{p+1}$ spacetime times a compact manifold, on one hand, and a conformal quantum field theory living on the stack of coincident D$p$-branes generating at low energies the above geometry, on the other hand.

Such correspondence was then very fruitfully expressed in quite general terms as a duality between AdS$_{d+1}$ supergravity and a conformal quantum field theory, CFT$_d$, living at its conformal boundary, and then further generalized as a gauge/gravity correspondence between a classical gravity theory (not necessarily with an AdS background) and a quantum field theory (not necessarily conformal) defined at its boundary.
It is a strong/weak coupling duality between
gravity at large (small) curvature radius and gauge theory at strong (weak) coupling, that implies a one-to-one correspondence between quantum operators $\mathcal{O}$ in the boundary field theory and fields $\phi$ of the bulk supergravity theory.  The formulation of the aforementioned duality, which then allowed to adapt tools from gravitational theories to explore non-perturbative properties of gauge theories and viceversa, requires to supplement the supergravity action functional with appropriate boundary conditions $\phi^{(0)}$ for the supergravity fields, which act as sources for the operators of the quantum field theory.
 
The literature on AdS/CFT and on its far reaching developments in various directions is so huge that we refrain from reviewing here further this seminal subject in general, and also from listing all the important references, which in any case could hardly be exhaustive. We limit ourselves here to refer to the first publications and to general reviews containing more extended reference lists. 

A holographic renormalization scheme was built on  \cite{Balasubramanian:1999re,deBoer:1999tgo,Verlinde:1999xm,deBoer:2000cz,deHaro:2000vlm,Skenderis:2002wp} to dispose of the divergences that show up in both sides of the correspondence. In particular, as far as the metric field is concerned, the bulk metric is divergent near the asymptotic boundary. However, these divergences can be successfully removed, in  the \textit{holographic renormalization} framework, through the inclusion of appropriate counterterms at the boundary.

The AdS/CFT duality is an important inspirational tool for the present review, but let us stress that we will actually focus on its relevance from the (super)gravity side of the correspondence.

In the geometrical  description of \cite{Aros:1999id,Aros:1999kt,Mora:2004kb,Olea:2005gb,Jatkar:2014npa,Jatkar:2015ffa,Miskovic:2009bm,Anastasiou:2020zwc}, using a Fefferman-Graham parametrization of spacetime  near the boundary, it was shown that the expansion   of the boundary Lagrangian in \eqref{EGBterm}, in the radial coordinate orthogonal to the boundary, is precisely the contribution needed to regularize the action and the related (background-independent) conserved charges. This contribution indeed reproduces the resummation (in an expansion in powers of the radial coordinate) of the holographic renormalization counterterms. The regularization involving the EGB term \eqref{EGBterm} can be compared to the holographic regularization procedure by adding and subtracting the York-Gibbons-Hawking term \cite{York:1972sj, Gibbons:1976ue} from the Einstein-Hilbert action plus a boundary term that is a specific polynomial in the extrinsic and intrinsic curvatures \cite{Miskovic:2009bm}. In particular, in four dimensions such boundary term, known as \textit{second Chern form} and referred to as \textit{Kounterterm} in \cite{Miskovic:2009bm}, emerges from the integration of the EGB term when applying Euler's theorem in the presence of a boundary (namely, considering the boundary formulation of topological invariants): the integration gives a term proportional to the Euler characteristic plus the integral of the second Chern form over the boundary.
Following the prescription above, one obtains the terms that define the Dirichlet problem in gravity plus a counterterm Lagrangian. Performing the asymptotic expansion of the latter one recovers the results of the holographic regularization scheme. The divergence cancellation provided by the counterterm series can be finally regarded as a ``\textit{topological regularization}''. Indeed, it is equivalent to adding the EGB term with a coupling such that the regularized action takes a MacDowell-Mansouri form \cite{MacDowell:1977jt}.

Let us remark that most of the work on the role of boundaries in supergravity, in the AdS/CFT context and also beyond it, mainly concerned the bosonic sectors of the given theories, while the inclusion of superpartners was not the subject of a large interest in the literature.
Some relevant contributions that are particularly worth mentioning, in various contexts and with different approaches, are \cite{Amsel:2009rr,vanNieuwenhuizen:2005kg,Esposito:1996hu,Avramidi:1997hy,Moss:2003bk,Moss:2004ck,Howe:2011tm}.

A systematic way to face (in principle) the boundary problem in supergravity was found in \cite{Andrianopoli:2014aqa} within the geometric approach to supergravity in superspace \cite{Castellani:1991et}, also known as the ``\textit{rheonomic approach}''.  This approach gives a geometric interpretation to the supersymmetry transformations rules, as diffeomorphisms in the fermionic directions of superspace. For a recent and nice review of this topic we refer the reader to \cite{DAuria:2020guc}. 

In \cite{Andrianopoli:2014aqa},  the supersymmetric Lagrangian in superspace in the presence of a non-trivial spacetime boundary was constructed, working in the geometric approach, for the four-dimensional $\mathcal{N} = 1$ and $\mathcal{N}=2$ pure supergravity theories with negative cosmological constant. It turned out that the supersymmetry invariance of the action is achieved with the inclusion of topological (boundary) terms corresponding to a supersymmetric extension of the EGB term \eqref{EGBterm}. 
Moreover, in this framework the field equations  involve non-trivial boundary contributions, and imply that   the boundary values of the super field-strengths are dynamically  fixed to constant values in the anholonomic basis of the bosonic and fermionic vielbein. The aforementioned conditions on the supercurvatures correspond to the vanishing of the $\rm{OSp}(1|4)$- and $\rm{OSp}(2|4)$-covariant super field-strengths at the boundary, respectively for the $\mathcal{N}=1$ and the $\mathcal{N}=2$ theory. The resulting supersymmetric Lagrangian  acquires a MacDowell-Mansouri form \cite{MacDowell:1977jt}, quadratic in the $\rm{OSp}(\mathcal{N}|4)$ supercurvatures.
 
Further applications of the superspace geometric approach to supergravity in the presence of a non-trivial boundary were subsequently presented in \cite{Ipinza:2016con,Banaudi:2018zmh}, where, however, models with an enlarged definition of supergravity and exhibiting a generalized cosmological constant \cite{deAzcarraga:2010sw,Soroka:2011tc,Durka:2011gm,Cebecioglu:2014rca,Concha:2015tla,Penafiel:2018vpe,Kibaroglu:2018oue}\footnote{With the terminology ``generalized cosmological constant'' we mean, according with the literature on the subject, a modification of the standard volume 4-form   in which a dependence on additional
1-form gauge  fields explicitly appears.} were considered. Let us also mention that recently the geometric approach to $\mathcal{N}=1$ and $\mathcal{N}=2$ supergravity theories with boundary was applied in \cite{Eder:2021nyb} in the context of loop quantum gravity to derive the so-called Holst-MacDowell-Mansouri action (involving, in particular, a parity odd term named Hojman term \cite{Hojman:1980kv,Iosifidis:2020dck}, and also known as Holst term \cite{Holst:1995pc}).

A general feature of all the models including boundaries that we have mentioned above is the presence in the bulk of a (negative) cosmological constant.
The limit case in which the cosmological constant vanishes  is a subtle issue, not explored  in \cite{Andrianopoli:2014aqa}.
This limit is sometimes  referred to as ``flat'' supergravity since the Lagrangian does not feature any explicit internal scale. 
Such case was first considered in \cite{Gibbons:1976ue}, and at the level of supergravity, including fermion contributions,
consistent boundary conditions in flat supergravity have been studied in \cite{Belyaev:2008ex,vanNieuwenhuizen:2005kg,vanNieuwenhuizen:2006pz}. 

In the geometric approach, a first result in this line of research was presented in \cite{Concha:2018ywv}, where the supersymmetry invariance of flat supergravity in four dimensions  with boundary was explored. The supersymmetry invariance of the
Lagrangian still requires to add appropriate boundary terms but, in this case, this is achieved by including additional 1-form fields which contribute only to the boundary Lagrangian. In the same paper it was shown that the supersymmetry-invariant theory found is  the vanishing cosmological constant limit of a particular deformation of AdS$_4$ supergravity with boundary involving a fields redefinition and  exhibiting a generalized cosmological constant.
Also in these cases, the resulting Lagrangian can be written in a MacDowell-Mansouri-like form in terms, as we will review in the following, of AdS-Lorentz supercurvatures in the theory with the generalized cosmological constant,
and
of the so-called minimal super-Maxwell curvatures \cite{Bonanos:2009wy,Concha:2014tca,Concha:2015tla} in its flat limit.  

From the physical point of view, the deformation of the supergravity theory  is functional to circumvent an obstruction in getting the flat supergravity as a limit of AdS$_4$ supergravity, in cases where spacetime has a non trivial boundary. In these situations, the straightforward zero cosmological constant limit of AdS$_4$ supergravity has a divergent boundary term. On the other hand, the deformation considered in \cite{Concha:2018ywv} corresponds to redefine spin connection and gravitino in the bulk AdS$_4$ supergravity by adding new tensor 1-forms, which vanish in the limit, but contribute, independently of the original fields,  to the boundary Lagrangian. This  allows to include in the boundary Lagrangian only those  contributions  that do not diverge in the flat limit.

At the algebraic level, the AdS-Lorentz supercurvatures are covariant with respect to a superalgebra corresponding to a supersymmetric extension of the AdS-Lorentz algebra \cite{Soroka:2006aj,Gomis:2009dm,Diaz:2012zza,Salgado:2013eut}, $\mathfrak{so}(D-1,1)\times\mathfrak{so}(D-1,2)$, which is a particular semi-simple extension of the Poincaré algebra involving the explicit presence of a scale parameter and an extra (with respect to the Poincaré symmetry) bosonic generator $Z_{\mu \nu}=-Z_{\nu \mu}$. Correspondingly, the AdS-Lorentz superalgebra is a semi-simple extension of the super-Poincaré algebra.

The Maxwell superalgebra on which the flat limit theory is based can be obtained by performing an In\"{o}n\"{u}-Wigner contraction on the length parameter of the AdS-Lorentz superalgebra.\footnote{The Maxwell algebra was firstly introduced to describe the symmetries of a particle
moving in a background in the presence of a constant electromagnetic field \cite{Schrader:1972zd}.} This is possible since also the Maxwell algebra involves an extra generator with respect to the Poincaré symmetry, namely $Z_{\mu \nu}$, which was originally associated with the electromagnetic field. Correspondingly, the minimal super-Maxwell curvatures we mentioned above correspond to a supersymmetric extension of the purely bosonic Maxwellian ones and can be obtained by performing an In\"{o}n\"{u}-Wigner contraction on the AdS-Lorentz supercurvatures. A peculiarity of the minimal supersymmetric extension of the Maxwell algebra is that its closure requires an extra odd generator, $\Sigma_\alpha$, which is nilpotent, besides the supercharge $Q_\alpha$,  dual to the gravitino 1-form field $\psi^\alpha$. This implies  the presence, in the dual formulation of the superalgebra in terms of Maurer-Cartan equations, of the corresponding   dual 1-form spinor field, $\chi^\alpha$.\footnote{In the superstring and supergravity contexts, superalgebras including nilpotent fermionic generators were  introduced in \cite{DAuria:1982uck} to consistently  formulate the hidden algebra underlying eleven dimensional supergravity,   and then analyzed in, e.g., \cite{Castellani:1982kd,Green:1989nn,Andrianopoli:2016osu,Andrianopoli:2017itj,Penafiel:2017wfr,Ravera:2018vra}.} 
Hence, with respect to the super-Poincaré algebra the super-Maxwell one is endowed with the extra even generators $Z_{ab}=-Z_{ba}$, written with Lorentz indices $a,b,\ldots=0,1,2,3$ in four spacetime dimensions, and odd generators $\Sigma_\alpha$, dual to the bosonic 1-form   field $A^{ab}=-A^{ba}$ and to the Majorana spinor 1-form $\chi^\alpha$, respectively.

For further details on the nowadays extended family of (super-)Maxwell algebras we refer the interested reader to \cite{Concha:2018ywv} and references therein. Here the crucial point will just be that the super-Maxwell curvatures can be obtained by performing an In\"{o}n\"{u}-Wigner contraction on the length parameter of the super AdS-Lorentz ones.

\vskip 5mm
The present review is essentially divided into two parts. In the first one we will deal with the geometric approach to the boundary problem in supergravity under a rather general perspective, independently of the choice of the boundary. With this we mean that we will consider the consequences of having non-vanishing fields on the boundary slice. However, we will not address in this part the question of where the boundary is located, nor whether it is space-like, time-like, or light-like. In the second part, we will show some applications of the results to specific problems, which require an explicit choice of the boundary. This issue also includes an amount of supersymmetry to be preserved on the boundary.

The remaining of this paper is structured as follows: In Section \ref{geombdy} we recall the key aspects of the geometric approach to supergravity in superspace in the presence of a non-trivial boundary of spacetime. In Section \ref{aads4} we review the geometric approach to the boundary problem in asymptotically AdS$_4$ supergravities, focusing, in particular, on the $\mathcal{N}=1$ and $\mathcal{N}=2$ pure supergravity theories. Section \ref{flat} is devoted to the boundary problem in four-dimensional supergravity in the absence of any internal scale in the Lagrangian, in which case supersymmetry invariance of the Lagrangian is recovered by supplementing the theory with appropriate boundary contributions given in terms of additional gauge fields which, however, do not appear in the bulk Lagrangian. Finally, the aforementioned theory is shown to properly emerge in the vanishing comological constant limit of a peculiar deformation, involving  fields redefintion, of AdS$_4$ supergravity.
Finally, in Section \ref{applicationssect} we report on applications of the results to specific problems.

\section{Geometric approach in the presence of a non-trivial boundary of spacetime}\label{geombdy}

Let us shortly review   the geometric approach to supergravity developed in \cite{Castellani:1991et},
restricting here  the analysis to the case of $D=4$ spacetime dimensions.

The geometric approach to supergravity \cite{Castellani:1991et} is a superspace approach. The theory is given in terms of 1-form superfields  $\mu^\mathcal{A}(x^\mu,\theta^{\alpha A})$ defined on $\mathcal{N}$-extended superspace $\mathcal{M}_{4|4\mathcal{N}} (x^\mu,\theta^{\alpha A})$, where  $x^\mu$ are commuting bosonic coordinates while $\theta^{\alpha A}$ are fermionic Grassmann coordinates ($\mu=0,1,2,3$ denotes spacetime indices, $\alpha=1,\ldots,4$ spinor indices, which we will generally omit in the following, and $A=1,\ldots,\mathcal{N}$ is an R-symmetry index labelling the number of supersymmetries).
Here, the index $\mathcal{A}$   collectively labels all the 1-forms  of the theory. These include
in particular the supervielbein $\lbrace V^a , \psi_A \rbrace$, which defines an orthonormal basis of superspace, $V^a$ being the bosonic vielbein ($a=0,1,2,3$ denotes anholonomic tangent space indices) and $\psi_A$ the $\mathcal{N}$-extended gravitino 1-form. Besides them, $\mu^\mathcal{A}$ also include the Lorentz spin connection $\omega^{ab}$ and all the internal symmetry 1-form gauge fields.
The set $\{\mu^\mathcal{A}\}$ defines the Maurer-Cartan 1-forms of the   theory, which encode the algebraic structure of the given supergravity theory through their Maurer-Cartan structure equations
\begin{equation}
    R^\mathcal{A}\equiv  \diff \mu^\mathcal{A} + \frac 12 {C^\mathcal{A}}_{\mathcal{B}\mathcal{C}}\mu^\mathcal{B} \wedge \mu^\mathcal{C} \,, \label{MCcurv}
\end{equation}
where ${C^\mathcal{A}}_{\mathcal{B}\mathcal{C}}$ are the structure constants of the supersymmetry algebra on which the supergravity theory considered is based. Here  $R^\mathcal{A}$ denote the supercurvature 2-forms, which are the field-strengths of the theory (also referred to as super field-strengths), whose vacuum value ($R^\mathcal{A}=0$) gives the superalgebra in its dual Maurer-Cartan formulation.

In the geometric superspace approach, supersymmetry transformations on spacetime are associated with diffeomorphisms in the fermionic ($\theta^{\alpha A}$) directions of superspace.
In this setting, supergravity theories are formulated from the condition of invariance under ``general super-coordinate transformations'', generalizing to superspace the geometric description of general relativity in terms of spacetime diffeomorphisms. 

In this setting, we denote by  $\mathcal{L}[\mu^\mathcal{A}]$ the (bosonic) Lagrangian   4-form in superspace and the action is obtained by integrating $\mathcal{L}$ on a generic bosonic hypersurface $\mathcal{M}_4 \subset \mathcal{M}_{4|4\mathcal{N}}$ immersed in superspace, namely
\begin{equation}
\mathcal{S} = \int _{\mathcal{M}_4 \subset \mathcal{M}_{4|4\mathcal{N}}} \mathcal{L}  [\mu^\mathcal{A}] \,.
\end{equation}
Indeed, in the geometric framework the Lagrangian is written in a background-independent (geometric) way, that is to say, independent of the choice of a metric. It is therefore invariant under general coordinate transformations in superspace (which include supersymmetry transformations on spacetime).\footnote{For this to be possible, the Lagrangian 4-form has to be entirely expressed as wedge product of differential forms and their exterior derivatives, without the use of tensor densities such as the Levi-Civita symbol $\epsilon_{\mu\nu\rho\sigma}$. For this reason, the kinetic terms in particular have to be written at first-order, thus avoiding the Hodge dual of the field-strengths, which is defined in terms of the Levi-Civita symbol.} Therefore, one can exploit general super-coordinate transformations to freely choose any $\mathcal{M}_4 \subset \mathcal{M}_{4|4\mathcal{N}}$ as the bosonic submanifold of integration in superspace, since any local deformation of the integration manifold can be reabsorbed in a superdiffeomorphism \cite{Castellani:1991et,DAuria:2020guc,Neeman:1978njh,Neeman:1978zvv}.

Let us stress that the superfield 1-forms $\mu^\mathcal{A}(x^\mu,\theta^{\alpha A})$, together with their field-strengths $R^\mathcal{A}(x^\mu,\theta^{\alpha A})$, are functions of all the coordinates of superspace,
\begin{equation}
    \mu^\mathcal{A} (x, \theta) = \mu^\mathcal{A}_\mu (x,\theta) \diff x^\mu + \mu^\mathcal{A}_{\alpha A} (x,\theta) \diff \theta^{\alpha A} \,,
\end{equation}
and they are related to the corresponding spacetime quantities $\mu^\mathcal{A}(x)=\mu^\mathcal{A}_\mu(x)\diff x^\mu$ by the restriction
\begin{equation}
    \mu^\mathcal{A}(x) = \mu^\mathcal{A} (x,\theta)|_{\theta=\diff \theta=0} = \mu^\mathcal{A}_\mu (x,0) \diff x^\mu \,.
\end{equation}
In principle, the theory defined in superspace could exhibit extra dynamics with respect to its spacetime projection. If this were the case, the resultant theory would fail to be equivalent to supergravity formulated in terms of a local spacetime (super)symmetry.
As we are going to discuss in the following, what allows the  formulation of supergravity in superspace to be equivalent to the one on spacetime  is the so-called rheonomy principle.

\subsection{The principle of rheonomy}\label{rheo}

In order for the theory on superspace to have the same physical content as the theory on spacetime, some constraints  have to be imposed on the supercurvatures (they were named ``rheonomic constraints'' in \cite{Castellani:1991et}, whence the name ``rheonomy" under which the geometric approach is also known). More precisely, in this framework the four-dimensional supergravity theories are formulated geometrically in terms of the supervielbein and of supercurvature 2-forms \eqref{MCcurv}, covariant under the superspace and internal symmetries of the theory. They extend to the supersymmetry representation of the given theory what in Einstein's theory is the Riemann tensor 2-form.

However, the supercurvatures can be actually expressed in two different ways, that have to be equivalent:
\begin{itemize}
\item They are defined off-shell from their symmetry properties, as in \eqref{MCcurv}, in terms of Lorentz and R-symmetry covariant exterior derivatives of the superfield 1-forms $\mu^\mathcal{A}$, and have to satisfy  consistency constraints given by the closure of Bianchi identities ($\diff^2=0$).
\item However, being 2-forms in superspace, they can also be expanded along the supervielbein basis $\lbrace V^a, \psi_A \rbrace$ of superspace,
\begin{equation}
    R^\mathcal{A} = {R^\mathcal{A}}_{ab}(x,\theta) V^a \wedge V^b + {R^\mathcal{A}}_{a \, \alpha A}(x,\theta) V^a \wedge \psi^{\alpha A} + {R^\mathcal{A}}_{\alpha A \, \beta B}(x,\theta) \psi^{\alpha A} \wedge \psi^{\beta B} \,, \label{param}
\end{equation}
where the superspace tensors ${R^\mathcal{A}}_{ab}$, appearing in the decomposition along  bosonic vielbein only, are referred to as ``inner components'', while the ones which appear in the decomposition along at least one fermionic direction, ${R^\mathcal{A}}_{a \, \alpha A}$ and ${R^\mathcal{A}}_{\alpha A \, \beta B}$, are the ``outer components''. Eq. \eqref{param} gives the so-called {\it{rheonomic parametrization}} of the supercurvatures, as they were defined in \cite{Castellani:1991et}.
\end{itemize}
The components in the parametrization of the supercurvatures can be determined by requiring that the supercurvatures satisfy  the corresponding Bianchi identities also when expressed in terms of their parametrizations. However,  this result can only be achieved on-shell, since the supersymmetry algebra only closes on-shell when represented in terms of dynamical fields.
One generally finds that the outer components  of the
supercurvatures have to be expressed, on-shell, as linear tensor combinations of the inner ones (which are actually known in the literature as \textit{supercovariant field-strengths}). These conditions are called {\it{rheonomic constraints}}. The same conclusion can be reached at the Lagrangian level, by decomposing the field equations with respect to independent sectors in  supervielbein polynomials in superspace.
From the physical point of view, the restriction on the superspace parametrizations given by the rheonomic constraints guarantees that no additional degrees of freedom are introduced in the theory in superspace compared to those already present in spacetime.

Let us briefly comment  on the general fact that the Bianchi identities of the theory in superspace are in fact not identities, but relations among the superfields and their curvatures, which are only satisfied \textit{on-shell}.
This property is a reflection of the fact that the Bianchi identities guarantee the closure of the given algebra when represented in terms of   fields. In this respect, the supersymmetry algebra is peculiar since supersymmetry representations have to contain the same number of bosonic and fermionic degrees of freedom (d.o.f. in the following). However,  the on-shell condition changes in different ways the number of d.o.f. of fields of different spin (e.g., spinors halve their d.o.f., while gauge vectors lower of one their d.o.f., and scalars do not change them). As a consequence, when supersymmetry is realized in terms of  field representations (supermultiplets), as it happens in supergravity theories,
it is an  on-shell symmetry.\footnote{This restriction can be relaxed by including auxiliary fields in the theory, but this is however possible at the supergravity level only in few cases (in theories exhibiting up to $8$ supercharges, and with the inclusion of an infinite number of auxiliary fields for the off-shell description of hypermultiplets). In those cases, the Bianchi identities are proper identities and the supersymmetry algebra closes off-shell.}

\vskip 5mm
Once the rheonomic constraints are exploited to find the parametrizations of all the supercurvatures, the latter then provide the supersymmetry transformation laws of the fields on spacetime, that leave invariant the spacetime Lagrangian  up to boundary terms. Indeed, as already mentioned, in the geometric framework the supersymmetry transformation laws of the fields in superspace can be read as diffeomorphisms generated by tangent vectors $\epsilon\equiv\bar\epsilon_{ A}Q^{ A}$ in the fermionic directions of superspace ($Q_{\alpha A}$ being the supersymmetry generators and $\epsilon_A$ an infinitesimal spinor  to be identified with the supersymmetry parameter),  and can therefore be expressed in terms of {\it{Lie derivatives}}.

Let us define $\imath_\epsilon$ as the contraction operator along  odd directions of superspace with parameter $\epsilon_A$, satisfying the property $\imath_\epsilon(\psi_A)=\epsilon_A$, $\imath_\epsilon(\mu^{\mathcal{A}})=0$ for $\mu^{\mathcal{A}}\neq \psi_A$. Then, we can write the supersymmetry transformation of a generic (tensor) superfield 1-form $\mathbf{\Phi}(x,\theta)$ as
\begin{equation}
    \delta_\epsilon \mathbf{\Phi}=\ell_\epsilon \mathbf{\Phi}= \imath_\epsilon (\diff \mathbf{\Phi}) + \diff(\imath_\epsilon \mathbf{\Phi})= \imath_\epsilon (\nabla \mathbf{\Phi}) + \nabla(\imath_\epsilon \mathbf{\Phi})\,,
\end{equation}
where $\nabla$ generically denotes covariant derivative with respect to the tensorial structure of the given superfield $\mathbf{\Phi}$. Note that, in performing the contraction, one has to use for $\nabla \mathbf{\Phi}$ its rheonomic parametrization as a 2-form in superpace, and not its formal definition as a covariant tensor under the symmetries of the given theory. This is a crucial requirement implementing, in the geometric approach, the condition that supersymmetry is realized on field representations as an on-shell symmetry, and then allowing the equivalence of diffeomorphisms in odd directions of superspace with supersymmetry transformations on spacetime. The same argument can be naturally generalized to the supersymmetry transformation of fields which are 0-forms and to higher-degree forms as well, by applying the principle of rheonomy on the same lines as for 1-forms.

For a complete presentation of the rheonomic approach to supergravity in superspace we refer the reader to the original source \cite{Castellani:1991et} and to the recent review \cite{DAuria:2020guc}.

\subsection{Supersymmetry invariance of the action}\label{susyinv}

The principal demand of any supergravity theory is the invariance of the action under supersymmetry  transformations,
\begin{equation}
    \delta_{\epsilon} \mathcal{S} \equiv  \int_{\mathcal{M}_4}\delta_\epsilon\mathcal{L}=0\,, \label{deltaSsusy}
\end{equation}
for $\epsilon\equiv\bar\epsilon_{ A}Q^{ A}$. Let us stress that in the geometric action the four-dimensional spacetime is consistently embedded as a four-dimensional hypersurface $\mathcal{M}_4 \subset \mathcal{M}_{4|4\mathcal{N}}$ 
on which the integration is performed. In fact, we can safely ignore the contribution from the integration manifold $\mathcal{M}_4$ when implementing the variational principle, since any variation of $\mathcal{M}_4$ can be compensated by a superdiffeomorphism under which the geometric Lagrangian is invariant \cite{Castellani:1991et,DAuria:2020guc,Neeman:1978njh,Neeman:1978zvv}. Supersymmetry transformations relate the Lagrangian on $\mathcal{M}_4$ to the one on any other submanifold $\mathcal{M}'_4$, entailing independence of the submanifold of integration in the action.

As we have just mentioned above, in the geometric approach the supersymmetry transformations are given by diffeomorphisms in the fermionic directions of superspace, so that the condition for the superspace Lagrangian 4-form to be invariant under local supersymmetry is
\begin{equation}\label{susyLagr}
    \delta_\epsilon \mathcal{L} = \ell_\epsilon \mathcal{L} = \imath_\epsilon (\diff \mathcal{L}) + \diff(\imath_\epsilon \mathcal{L}) = 0 \,.
\end{equation}
Notice that the first contribution in \eqref{susyLagr}, which would be identically zero in spacetime, is instead non-trivial here, since $\mathcal{L}$ is not a top form in $\mathcal{N}$-extended superspace (which has dimension $4+4\mathcal{N}$), so that the 5-form $\diff \mathcal{L}$ is non-vanishing in superspace. The second contribution in \eqref{susyLagr} is a boundary term that does not affect the bulk result, and that can be discarded in the absence of non-trivial boundaries, due to the general convergence conditions imposed in that case on the fields at spatial infinity. Hence, a necessary condition for a supersymmetry-invariant supergravity Lagrangian is
\begin{equation}\label{necessary}
    \imath_\epsilon (\diff \mathcal{L}) = 0 \,,
\end{equation}
corresponding to requiring supersymmetry invariance in the bulk of superspace. From now on we assume true the condition \eqref{necessary}, and the Lagrangians satisfying it  will be referred to as {\it{bulk supergravity Lagrangians}}, $\mathcal{L}_{\text{bulk}}$.
On the other hand, supersymmetry invariance of the action then  requires the weaker condition on the bulk Lagrangian
\begin{equation}
    \delta_\epsilon \mathcal{S} = \int_{\mathcal{M}_4} \diff (\imath_\epsilon \mathcal{L}_{\text{bulk}}) = \int_{\partial \mathcal{M}_4} \imath_\epsilon \mathcal{L}_{\text{bulk}} = 0 \,,
\end{equation}
that is
\begin{equation}\label{cond}
    \imath_\epsilon \mathcal{L}_{\text{bulk}} |_{\partial \mathcal{M}_4} = \diff \varphi \,,
\end{equation}
where $\partial \mathcal{M}_4$ is the boundary of $\mathcal{M}_4$ and where $\diff \varphi$ here is used to denote an exact quantity. However, eq. \eqref{cond} is in general not satisfied by $\mathcal{L}_{\text{bulk}}$ in the presence of  non-trivial   boundary conditions on the boundary $\partial \mathcal{M}_4$. In this case, supersymmetry invariance requires to add topological (boundary) terms. Thus, one has to consider the full Lagrangian
\begin{equation}
     \mathcal{L}_{\text{full}} = \mathcal{L}_{\text{bulk}} + \mathcal{L}_{\text{bdy}} \,,
\end{equation}
where $\mathcal{L}_{\text{bdy}}$ is given by  boundary contributions, which do not affect $\mathcal{L}_{\text{bulk}}$:
\begin{equation}
\mathcal{L}_{\text{bdy}}= \diff \mathcal{B}_{(3)}\,, \label{db}
\end{equation}
$\mathcal{B}_{(3)}$ denoting a 3-form.
Due to the form of the boundary Lagrangian, \eqref{db}, we immediately see that $\imath_\epsilon (\diff \mathcal{L}_{\text{full}}) = 0$. But the crucial point is that the inclusion of such boundary contribution allow to reestablish the supersymmetry invariance of the total Lagrangian, besides  introducing a boundary dynamics. In particular, as we will discuss in the following, the condition 
\begin{equation}
    \imath_\epsilon \mathcal{L}_{\text{full}} |_{\partial \mathcal{M}_4} = 0
\end{equation}
(up to an irrelevant total derivative $\diff \varphi$) determines the coefficients appearing in $\mathcal{L}_{\text{bdy}}$. Furthermore, the boundary contributions to the field equations\footnote{Typically, besides the contributions to the equations of motion coming from $\mathcal{L}_{\text{bdy}}$ we also have extra
contributions from $\mathcal{L}_{\text{bulk}}$, neglected in the absence of a boundary, originating from the total differentials obtained by partial integration.} of the theory on $\mathcal{M}_{4|4\mathcal{N}}$ fix in a dynamical way the supercurvatures on $\partial \mathcal{M}_4$ to constant values in the anholonomic basis of the bosonic and fermionic vielbein.

\section{Facing the boundary problem in AdS$_4$ pure supergravities}\label{aads4}

In this section, following \cite{Andrianopoli:2014aqa}, we review the geometric construction of the four-dimensional $\mathcal{N}=1$ and $\mathcal{N}=2$ pure supergravity Lagrangians with negative cosmological constant in the presence of a non-trivial boundary of spacetime. These are the simplest theories that can be considered, since for these cases the supergravity multiplets do not include fields of spin lower than $1$. This implies that the theory can be formulated fully geometrically only in terms of 1-form fields and their supercurvatures.

As we are going to review, the  boundary terms needed to recover supersymmetry
invariance of the action can be interpreted as the $\mathcal{N}$-supersymmetric extension of the  Euler-Gauss-Bonnet term.

\subsection{$\mathcal{N}=1$ case}\label{N1}

The field content of the $\mathcal{N}=1$, $D=4$ pure supergravity theory is given by the bosonic vielbein 1-form $V^a$, the Lorentz spin connection $\omega^{ab}$ and one gravitino 1-form $\psi^\alpha$, which is a Majorana spinor spanning the fermionic directions of $\mathcal{N}=1$ superspace. Here and in the following sections, our convention for the metric signature is mostly minus.

Since pure $\mathcal{N}=1$ supergravity is the simplest supersymmetric extension of Einstein's general relativity, we will use this model as an example of application of the geometric approach to supergravity discussed in Section \ref{rheo}, also holding beyond the specific focus on the boundary problem of the present review. 

The Lorentz-covariant  supercurvatures  of the theory are defined as follows:\footnote{Regarding our notation, with respect to \cite{Andrianopoli:2014aqa} here we define the Lorentz spin connection and curvature with extra minus signs, that is $\omega^{ab} \rightarrow - \omega^{ab}$ and $\mathcal{R}^{ab} \rightarrow - \mathcal{R}^{ab}$, to better fit in the conventions more commonly adopted in the literature.}
\begin{equation}\label{N1supercurv}
\begin{split}
\mathcal{R}^{ab} & \equiv \diff \omega^{ab} + {\omega^a}_c \wedge \omega^{cb} \,, \\
R^a & \equiv \mathcal{D} V^a - \frac{\ii}{2} \bar{\psi} \gamma^a \wedge \psi = \diff V^a + {\omega^{a}}_b \wedge V^b - \frac{\ii}{2} \bar{\psi} \gamma^a \wedge \psi \,, \\
\rho & \equiv \mathcal{D} \psi = \diff \psi + \frac{1}{4} \omega^{ab} \gamma_{ab} \wedge \psi \,,
\end{split}
\end{equation}
where $\mathcal{R}^{ab}$ is the Lorentz supercurvature 2-form,   $R^a$ is the supertorsion, $\rho$ is the gravitino super field-strength, $\mathcal{D}\equiv \diff +\omega$ denotes the Lorentz-covariant differential operator (acting differently in the fundamental and in the spinorial representation), while $\gamma_a$ and $\gamma_{ab}=\gamma_{[a}\gamma_{b]}$ are gamma matrices in four dimensions.\footnote{We adopt the same gamma matrices conventions of \cite{Andrianopoli:2014aqa}. For useful formulas on gamma matrices and spinor identities in $\mathcal{N}=1$ we refer the reader to the Appendix of \cite{DAuria:2021dth}, where some typos appearing in \cite{Andrianopoli:2014aqa} have also been fixed.} 
In the following, to lighten the notation we will generally omit  writing  the wedge product between differential forms.

The requirement that the supercurvatures \eqref{N1supercurv} satisfy (on-shell) the Bianchi identities
\begin{equation}\label{BIN1}
    \begin{split}
        \mathcal{D} \mathcal{R}^{ab} & = 0 \,, \\
        \mathcal{D} \rho & = \frac{1}{4} \mathcal{R}^{ab} \gamma_{ab} \psi \,, \\
        \mathcal{D} R^a & = {\mathcal{R}^a}_b V^b + \ii \bar{\psi} \gamma^a \rho \,
    \end{split}
\end{equation}
determines the  general expression of their parametrizations to be
\begin{eqnarray}
\left\{\begin{array}{lll}
\mathcal R^{ab}&=& {\mathcal{R}^{ab}}_{cd} V^c V^d  + \ii \left( \bar\rho^{ab} \gamma_c-2\bar\rho^{[a}_{\phantom{[a}c}\gamma^{b]}\right) \psi V^c +\frac 1{2\ell} \bar \psi \gamma^{ab}\psi \,, \\
\rho&=& \rho_{ab} V^a V^b + \frac\ii{2\ell} \gamma_a \psi V^a \,, \\
R^a &=& 0 \,, \end{array}\right. \label{N1param}
\end{eqnarray}
where the tensors ${\mathcal{R}^{ab}}_{cd}$ and $\rho_{ab}$ defining the components of the supercurvatures along two bosonic vielbein are the supercovariant field-strengths.
Note that, as it was first shown in \cite{Townsend:1977qa}, the paramerizations \eqref{N1param} allow the inclusion of terms proportional to a parameter  $\ell$  (a Fayet-Iliopoulos term \cite{Fayet:1974jb}) with the dimensions of a length (in natural unites). As we are going to see, the parameter $\ell$ can be interpreted as radius of AdS$_4$ background spacetime, while the limit $\ell\to \infty$  describes the supergravity theory on a Minkowski background.\footnote{In \cite{Andrianopoli:2014aqa}, the AdS$_4$ radius and  the cosmological constant are expressed in terms of the parameter $e =\frac{1}{2 \ell}$.}
In terms of the supercurvatures \eqref{N1supercurv} and of their parametrizations \eqref{N1param}, following the prescriptions in Section \ref{rheo}, one can determine the supersymmetry transformations of the fields in the supergravity multiplet to be
\begin{eqnarray}
\left\{\begin{array}{lll}
\delta_\epsilon \omega^{ab}&=& \ii \left( \bar\rho^{ab} \gamma_c-2\bar\rho^{[a}_{\phantom{[a}c}\gamma^{b]}\right) \epsilon V^c +\frac 1{\ell} \bar \epsilon \gamma^{ab}\psi \,, \\
\delta_\epsilon \psi&=& \mathcal{D}\epsilon + \frac\ii{2\ell} \gamma_a \epsilon V^a \,, \\
\delta_\epsilon V^a &=&\ii \bar \epsilon \gamma^{a}\psi \,. \end{array}\right. \label{N1susytransf}
\end{eqnarray}

We now consider the bulk Lagrangian of the theory in superspace, whose equations of motion admit an AdS$_4$ vacuum solution with negative cosmological constant $\Lambda = - \frac{3}{\ell^2}$. 
The bulk Lagrangian 4-form reads
\begin{equation}\label{LbulkN1}
    \mathcal{L}^{\mathcal{N}=1}_{\text{bulk}} = \frac{1}{4} \mathcal{R}^{ab} V^c V^d \epsilon_{abcd} - \bar{\psi} \gamma_5 \gamma_a \rho V^a - \frac{\ii}{2 \ell} \bar{\psi} \gamma_5 \gamma_{ab} \psi V^a V^b - \frac{1}{8 \ell^2} V^a V^b V^c V^d \epsilon_{abcd} \,,
\end{equation}
where $\epsilon_{abcd}$ is the four-dimensional Levi-Civita tensor.
It is written in first-order formalism for the spin connection $\omega^{ab}$, whose field equation enforces (up to boundary terms, which will be considered in a while) the vanishing  of the supertorsion $R^a$ defined in \eqref{N1supercurv}, as equivalently found in \eqref{N1param} by solving the Bianchi relations.

The Lagrangian \eqref{LbulkN1} is off-shell invariant under supersymmetry, since the condition \eqref{necessary} holds,
\begin{equation}
    \imath_\epsilon (\diff \mathcal{L}^{\mathcal{N}=1}_{\text{bulk}}) = 0 \,,\label{nec1}
\end{equation}
as it can be explicitly checked using \eqref{N1supercurv} to evaluate $\diff \mathcal{L}^{\mathcal{N}=1}_{\text{bulk}}$ off-shell, and then  \eqref{N1param} to perform the contraction of the supercurvatures in \eqref{nec1} along the odd directions  $\epsilon \equiv \bar\epsilon Q$ of superspace.

On the other hand, as discussed in Section \ref{susyinv}, when the background spacetime has a non-trivial boundary, the condition $\imath_\epsilon \mathcal{L} |_{\partial \mathcal{M}_4} = 0$ (modulo an exact differential) is in general not satisfied, and it is necessary to check it explicitly to get supersymmetry invariance of the action. In fact, in the case at hand we find that, without imposing  boundary conditions on the fields,
\begin{equation}
    \imath_\epsilon \mathcal{L}^{\mathcal{N}=1}_{\text{bulk}} |_{\partial {\mathcal{M}_4}} \neq \diff \varphi \quad \Rightarrow \quad \delta_\epsilon \mathcal{S}_{\text{bulk}} \neq 0 \,.
\end{equation}
As shown in \cite{Andrianopoli:2014aqa}, the supersymmetry invariance of the theory is restored  by adding appropriate boundary terms $\mathcal{L}^{\mathcal{N}=1}_{\text{bdy}}=\diff \mathcal{B}_{(3)}$ to the superspace Lagrangian
\begin{equation}
\mathcal{L}^{\mathcal{N}=1}_{\text{bulk}} \quad \rightarrow \quad \mathcal{L}^{\mathcal{N}=1}_{\text{full}} \equiv\mathcal{L}^{\mathcal{N}=1}_{\text{bulk}}+\mathcal{L}^{\mathcal{N}=1}_{\text{bdy}}\,,  
\end{equation}
which do not alter $\diff \mathcal{L}^{\mathcal{N}=1}_{\text{bulk}}$ so that still $\imath_\epsilon (\diff \mathcal{L}^{\mathcal{N}=1}_{\text{full}}) = 0$.
Here, the possible boundary terms, compatible with the  parity and Lorentz invariances of the theory, are 
\begin{equation}\label{bdyterms1}
    \begin{split}
      & \diff \left(\omega^{ab} \wedge \mathcal{R}^{cd} + {\omega^a}_{{f}} \wedge \omega^{{f}b} \wedge \omega^{cd} \right) \epsilon_{abcd} = \mathcal{R}^{ab} \wedge \mathcal{R}^{cd} \epsilon_{abcd} \,, \\
      & \diff \left( \bar{\psi} \wedge \gamma_5 \rho \right) = \bar{\rho} \wedge \gamma_5 \rho - \frac{1}{4} \mathcal{R}^{ab} \wedge \bar{\psi} \gamma_5 \gamma_{ab} \wedge \psi \,.
    \end{split}
\end{equation}
We therefore consider the boundary Lagrangian
\begin{equation}\label{LbdyN1}
    \mathcal{L}^{\mathcal{N}=1}_{\text{bdy}} = \alpha \mathcal{R}^{ab} \mathcal{R}^{cd} \epsilon_{abcd} -\ii \beta \left( \bar{\rho} \gamma_5 \rho - \frac{1}{4} \mathcal{R}^{ab} \bar{\psi} \gamma_5 \gamma_{ab} \psi  \right) \,,
\end{equation}
where the coefficients $\alpha$ and $\beta$ are real parameters to be determined by the request of supersymmetry invariance of the full Lagrangian
\begin{equation}\label{LfullN1}
   \begin{split}
       \mathcal{L}^{\mathcal{N}=1}_{\text{full}} 
       & = \frac{1}{4} \mathcal{R}^{ab} V^c V^d \epsilon_{abcd} - \bar{\psi} \gamma_5 \gamma_a \rho V^a - \frac{\ii}{2 \ell} \bar{\psi} \gamma_5 \gamma_{ab} \psi V^a V^b - \frac{1}{8 \ell^2} V^a V^b V^c V^d \epsilon_{abcd} \\
       & + \alpha \mathcal{R}^{ab} \mathcal{R}^{cd} \epsilon_{abcd} - \ii \beta \left( \bar{\rho} \gamma_5 \rho - \frac{1}{4} \mathcal{R}^{ab} \bar{\psi} \gamma_5 \gamma_{ab} \psi  \right) \,,
   \end{split}
\end{equation}
that is by imposing the condition
\begin{equation}
    \left[\diff \left(\imath_\epsilon (\mathcal{L}^{\mathcal{N}=1}_{\text{full}})\right)\right]_{ \mathcal{M}_4} = 0 \quad \Rightarrow \quad \left[\imath_\epsilon (\mathcal{L}^{\mathcal{N}=1}_{\text{full}})\right]_{\partial \mathcal{M}_4} = \diff \varphi\,.
\end{equation}

We should now consider the boundary contributions in the field equations from the full Lagragian \eqref{LfullN1}, which  result in the following constraints on the supercurvatures, to hold on the boundary:
\begin{equation}\label{curvbdyN1}
\begin{cases}
\frac{\delta \mathcal{L}^{\mathcal{N}=1}_{\text{full}}}{\delta \omega^{ab}} = 0 \quad \Rightarrow & \mathcal{R}^{ab} |_{\partial \mathcal{M}_4} = - \frac{1}{8 \alpha} \left( V^a V^b + \frac{1}{2} \beta \bar{\psi} \gamma^{ab} \psi \right)_{\partial \mathcal{M}_4} \,, \\
\frac{\delta \mathcal{L}^{\mathcal{N}=1}_{\text{full}}}{\delta \psi} = 0 \quad \Rightarrow & \rho |_{\partial \mathcal{M}_4} = \frac{\ii}{2 \beta} \left( \gamma_a \psi V^a \right)_{\partial \mathcal{M}_4} \,.
\end{cases}
\end{equation}
The above conditions imply that both the supercurvatures $\mathcal{R}^{ab}$ and $\rho$ on $\partial \mathcal{M}_4$ are not propagating, but rather dynamically fixed to constant values in the anholonomic basis of the bosonic and fermionic vielbeins. 

Finally, we have to impose supersymmetry invariance. Upon use of \eqref{curvbdyN1}, we find
\begin{equation}
\imath_\epsilon (\mathcal{L}^{\mathcal{N}=1}_{\text{full}} )|_{\partial \mathcal{M}_4} = 0 \quad \Leftrightarrow  \quad  \frac{\beta}{16 \alpha} - \frac{1}{2 \beta} = -\frac{1}{\ell} \,. 
\end{equation}
Solving the latter for $\beta$ (with the condition $\beta \neq 0$) we get
\begin{equation}
\beta = - \frac{8}{\ell} \alpha (1+k) \,, \quad \text{where}  \quad k =\pm \sqrt{ 1 + \frac{\ell^2}{8 \alpha}}\in \mathbb{R} \,.
\end{equation}
The above relations can be solved in terms of the real parameter $k\neq -1$, to give
\begin{equation}
\alpha =- \frac 18 \frac{\ell^2}{1-k^2}\,,\quad \beta =\frac{\ell}{1-k}\,,   
\label{paramN1}\end{equation}
that is
\begin{equation}\label{curvbdyN1bis}
\begin{cases}
 & \mathcal{R}^{ab} |_{\partial \mathcal{M}_4} = \left[ \frac{1-k^2}{\ell^2}V^a V^b + \frac{ 1+k }{2\ell} \bar{\psi} \gamma^{ab} \psi \right]_{\partial \mathcal{M}_4} \,, \\
 & \rho |_{\partial \mathcal{M}_4} = \frac{{\ii}(1-k)}{2 \ell} \left[ \gamma_a \psi V^a \right]_{\partial \mathcal{M}_4} \,.
\end{cases}
\end{equation}
Let us remark that the boundary terms \eqref{LbdyN1}, with parameters satisfying \eqref{paramN1}, are the $\mathcal{N}=1$ supersymmetric extension of the Euler-Gauss-Bonnet term.

Interestingly, setting $k=0$, which implies $\alpha=-\frac{\ell^2}{8}$ and $\beta=\ell$, the full Lagrangian \eqref{LfullN1} takes the form\footnote{\label{hodge}In writing the expression above we are skipping some subtleties related to the fact that the extension of the action integral to superspace requires, properly speaking, that the 4-form Lagrangian in superspace should be written at first-order, thus avoiding use of the Hodge-dual which is not easily defined in superspace  (unless one uses the formalism of integral forms in superspace, see \cite{Castellani:2014goa,Castellani:2015paa}). However, these subtleties do not affect the arguments reviewed here. }
\begin{equation}\label{MMN1}
    \mathcal{L}^{\mathcal{N}=1}_{\text{full}} =  - \frac{\ell^2}{8}  {\mathbf{R}}^{ab} \wedge  {\mathbf{R}}^{cd} \epsilon_{abcd} - \ii \ell  {\bar{\boldsymbol{\rho}}} \gamma_5 \wedge {\boldsymbol{\rho}} \,,
\end{equation}
which is written in terms of the $\rm{OSp}(1|4)$-covariant supercurvatures
\begin{equation}\label{osp14curv}
    \begin{split}
        {\mathbf{R}}^{ab} & \equiv \mathcal{R}^{ab} - \frac{1}{\ell^2} V^a V^b  - \frac{1}{2 \ell} \bar{\psi} \gamma^{ab} \psi \,, \\  {\boldsymbol{\rho}} & \equiv \rho - \frac{\ii}{2 \ell} \gamma_a \psi V^a \,, \\
   {\mathbf{R}}^{a} & \equiv   R^a\,,
    \end{split}
\end{equation}
 vanishing on the boundary due to \eqref{curvbdyN1}.
 
When written in the form \eqref{MMN1} holding for $k=0$, the Lagrangian $\mathcal{L}^{\mathcal{N}=1}_{\text{full}}$ can be recognized to be the MacDowell-Mansouri Lagrangian \cite{MacDowell:1977jt}, which is quadratic in the $\rm{OSp}(1|4)$-covariant super field-strengths.\footnote{Here let us emphasize that, in complete analogy to what has been observed in \cite{MacDowell:1977jt} for the case of AdS$_4$ gravity, the action \eqref{MMN1} is not invariant under local $\rm{OSp}(1|4)$ transformations, even though the supercurvatures \eqref{osp14curv} are covariant with respect to $\rm{OSp}(1|4)$.}
In fact, it corresponds the $\mathcal{N}=1$ supersymmetric extension of what was found for AdS$_4$ gravity in \cite{Miskovic:2009bm}, where the topologically renormalized action including the Euler-Gauss-Bonnet term was cast in the MacDowell-Mansouri form \cite{MacDowell:1977jt}. 

Note that, in terms of the $\rm{OSp}(1|4)$ supercurvatures \eqref{osp14curv}, the constraints \eqref{curvbdyN1} coming from the boundary contributions in the  field equations  take the simple form
\begin{equation}\label{vanosp14c}
  {\mathbf{R}}^{ab}|_{\partial \mathcal{M}_4} = 0 \,, \quad {\boldsymbol{\rho}}|_{\partial \mathcal{M}_4} = 0 \,, \quad {\mathbf{R}}^a|_{\partial \mathcal{M}_4}=0 \,, \qquad \text{for} \quad k=0 \,,
\end{equation}
where the last relation follows because the supertorsion $R^a$ vanishes on-shell  in the  whole superspace and then in particular, for continuity,  we also have ${\mathbf{R}}^a|_{\partial\mathcal M}=0$.
Note that equations \eqref{vanosp14c} imply that, for the case $k=0$, the fields of the theory satisfy the Maurer-Cartan equations of the $\rm{OSp}(1|4)$ algebra, that is the boundary in this case enjoys global invariance under $\rm{OSp}(1|4)$ symmetry. 
This is the $\mathcal{N}=1$ supersymmetric extension of the results in \cite{Aros:1999id,Aros:1999kt,Mora:2004kb,Olea:2005gb,Jatkar:2014npa,Jatkar:2015ffa}, following from the request  of invariance of the gravity Lagrangian under spacetime diffeomorphisms.

Let us emphasize that the same supergroup $\rm{OSp}(1|4)$ is also the $\mathcal{N}=1$ superconformal symmetry of $\mathcal{N}=1$  superspace in three dimensions, which allows for a natural holographic interpretation of the above result in a holographic context.

However, the $\mathcal{N} = 1$ supergravity theory we have discussed also allows for $k \neq 0$, in which case the MacDowell-Mansouri Lagrangian above is supplemented by extra boundary terms breaking the  $\rm{OSp}(1|4)$ structure of the theory. This freedom is peculiar of the minimal theory, and can be traced back to the fact that $\mathcal{N}=1$ is the least supersymmetric theory, and therefore the least constrained by supersymmetry in four dimensions.
On the contrary, as we are going to see in the following, in the $\mathcal{N}=2$ case all the coefficients of the boundary terms needed to restore supersymmetry invariance will result to be fixed to the values corresponding to rigid superconformal invariance of the boundary theory.

\subsection{$\mathcal{N}=2$ case}\label{N2}

Now that we have taken some familiarity with the geometric approach to the boundary problem in the case of $\mathcal{N}=1$, pure AdS$_4$ supergravity, let us move on, still following \cite{Andrianopoli:2014aqa}, to the slightly more complicated case of $\mathcal{N}=2$, $D=4$ pure supergravity with negative cosmological constant $\Lambda=-\frac{3}{\ell^2}$ in the presence of a non-trivial boundary of spacetime. In this case, as we shall see in the following, all the coefficients of the boundary terms will be fixed and no terms breaking the global boundary invariance $\rm{OSp}(2|4)$ are allowed by supersymmetry invariance.

The field content of the theory is given by the bosonic vielbein 1-form $V^a$, the gravitini 1-forms $\psi_{A}$, which are Majorana spinors (the index $A=1,2$ is in the fundamental representation of the R-symmetry group), the $\rm{SO}(1,3)$ spin connection $\omega^{ab}$, and the graviphoton $A^0$, which is an Abelian gauge connection, and in the following will be simply denoted by $A$. 

The bulk theory on AdS background can be obtained from the general $\mathcal{N} = 2$ matter coupled supergravity theory of \cite{Andrianopoli:1996cm} by setting to zero the gauge supermultiplets (so that the index $\mathbf{\Lambda}$ enumerating the gauge vectors only takes the value $\mathbf{\Lambda} =0$ corresponding to the graviphoton), and also the hypermultiplets, while keeping, however, a non-vanishing Fayet-Iliopoulos term $P$, proportional to the inverse of the AdS$_4$ radius $\ell$, out of the $\rm{SU}(2)$-valued momentum maps ${P}^x_{\mathbf{\Lambda}=0}$ ($x=1,2,3$) in the hypermultiplet sector.\footnote{Here, as for the $\mathcal{N}=1$ case, we shall adopt the notation of \cite{Andrianopoli:2014aqa}, but performing the changes $\omega^{ab}\rightarrow -\omega^{ab}$, $\mathcal{R}^{ab}\rightarrow -\mathcal{R}^{ab}$, and $A \rightarrow -\frac{1}{\sqrt{2}} A$.  Our conventions on fermions can be found in Appendix A.2 of \cite{Andrianopoli:2020zbl}, where the notation of \cite{Andrianopoli:2019sip} was adopted. We generally use Majorana spinors, and redefine the constants appearing in \cite{Andrianopoli:2014aqa} as $L^0=L=\frac{1}{\sqrt 2}$ and
$\frac 1\ell=2e=\frac{P}{\sqrt 2}=\sqrt{ -\frac{\Lambda}{3}}$.}
More precisely, the gravitino mass matrix reduces to the term
 $S_{AB}= \frac \ii 2 (\sigma^x)_A{}^{C} \epsilon_{CB} P^x_0 L^0 =  PL^0 \delta_{AB}= \frac 1\ell \delta_{AB}$.  
The choice of the Fayet-Iliopoulos term $P$, as pointing in a given fixed $\rm{SU}(2)$ direction, breaks
the R-symmetry, which would be $\rm{U}(2)$ for the ungauged theory, to $\mathrm{SO}(2)$ for AdS$_4$ supergravity.

The bulk Lagrangian 4-form in $\mathcal{N} = 2$ superspace is \cite{Andrianopoli:2014aqa,Andrianopoli:1996cm}
\begin{equation}\label{LbulkN2}
\begin{split}
\mathcal{L}^{\mathcal{N}=2}_{\text{bulk}} & =  \frac{1}{4} \mathcal{R}^{ab}V^cV^d\epsilon_{abcd}+\bar \psi ^A\gamma_a\gamma_5 \rho_{A}V^a+\frac{\ii}{2}\left( F+\frac{1}{2} \bar \psi^A \psi^B\epsilon_{AB}\right)\bar\psi ^C\gamma_5\psi^D\epsilon_{CD} \\
&-\frac{\ii}{2\ell}\bar\psi^A\gamma_{ab}\gamma_5\psi_{A}V^aV^b-\frac{1}{8\ell^2}V^aV^bV^cV^d\epsilon_{abcd} \\ 
&+\frac{1}{4}\left(\tilde F^{cd}V^aV^b F-\frac{1}{12}\tilde F_{lm} \tilde F^{lm}V^aV^bV^cV^d\right)\epsilon_{abcd} \,,
\end{split}
\end{equation}
which is written at first order  for the spin connection $\omega^{ab}$ and for the gauge field $A$. In particular, from the variation of \eqref{LbulkN2} with respect to the tensor 0-form $\tilde{F}^{cd}$ we get the condition $\tilde{F}_{ab}=F_{ab}$, $F_{ab}$ being the component along the purely bosonic vielbein of the field-strength 2-form $F$, that is the corresponding supercovariant field-strength. On the other hand, the field equation of the spin connection constrains the supertorsion
\begin{equation}
     {R}^{a} \equiv \mathcal{D} V^a -\frac{\ii}{2}\bar\psi^A \gamma^a \psi_A 
\end{equation}
to vanish on-shell.
Finally, the superspace 2-form supercurvatures appearing in \eqref{LbulkN2}
are defined  by
\begin{equation}\label{curvN2}
    \begin{split}
    \mathcal{R}^{ab}& \equiv \diff \omega^{ab} + \omega^{ac} {\omega_c}^b \,, \\
    \rho_A & \equiv \mathcal{D} \psi_A - \frac{1}{2\ell} A \epsilon_{AB} \psi^B = \diff \psi_A +\frac{1}{4} \omega^{ab}\gamma_{ab} \psi_A- \frac{1}{2\ell} A \epsilon_{AB} \psi^B \,, \\
    F & \equiv  \diff A - \bar \psi^A \psi^B\epsilon_{AB} \,.
    \end{split}
\end{equation}

Analogously to the previously discussed $\mathcal{N}=1$ case, a consistent definition of the supersymmetric action in the presence of a non-trivial spacetime boundary requires the full Lagrangian to include a boundary contribution, namely
\begin{equation}
    \mathcal{L}^{\mathcal{N}=2}_{\text{full}} = \mathcal{L}^{\mathcal{N}=2}_{\text{bulk}} + \mathcal{L}^{\mathcal{N}=2}_{\text{bdy}} \,,
\end{equation}
where, for the case at hand, following the same procedure we have reviewed for the $\mathcal{N}=1$ theory, the boundary Lagrangian is now given by
\begin{equation}\label{LbdyN2}
\mathcal{L}^{\mathcal{N}=2}_{\text{bdy}} =  \frac 14
     \diff \lbrace \alpha (\omega^{ab} \mathcal{R}^{cd} - {\omega^{a}}_{{f}}  \omega^{{f} b}  \omega^{cd})\epsilon_{abcd}  +\ii \beta \left(S_{AB} \bar\psi^A \rho^B -{\bar S}^{AB} {\bar\psi}_A \rho_B\right) +
\theta A\mathcal{F} \rbrace \,,
\end{equation}
where the $\mathcal{N}=2$ supersymmetry invariance determines  the coefficients to be
\begin{equation}
\alpha= \frac {\ell^2}2\,,\quad \beta =  8{\ell^2}\,,\quad \theta =0\,.
\end{equation}
These are the values reproducing as the boundary Lagrangian the $\mathcal{N}=2$ supersymmetric generalization of the purely bosonic Euler-Gauss-Bonnet term, that is
\begin{equation}\label{LbdyN2EGB}
\mathcal{L}^{\mathcal{N}=2}_{\text{bdy}} = -\frac{\ell^2}{8}\left(\mathcal R^{ab} \mathcal R^{cd}\epsilon_{abcd}+\frac{8{\ii}}{\ell} \bar{\rho}^A\gamma_5 \rho_{A}-\frac{2 \ii}{\ell}\mathcal R^{ab}\bar\psi^A\gamma_{ab}\gamma_5\psi_{A}+\frac{4 \ii}{\ell^2}\diff A \bar\psi^A\gamma_5\psi^B\epsilon_{AB}\right) \,.
\end{equation}
Let us mention in particular that the supersymmetry invariance of $\mathcal{L}^{\mathcal{N}=2}_{\text{full}}$ does not allow for a theta-term in the gauge sector. 

The supersymmetric full Lagrangian then reads
\begin{equation}\label{LfullN2}
    \begin{split}
        \mathcal{L}^{\mathcal{N}=2}_{\text{full}} & = \frac{1}{4} \mathcal{R}^{ab}V^cV^d\epsilon_{abcd}+\bar \psi ^A\gamma_a\gamma_5 \rho_{A}V^a+\frac{\ii}{2}\left( F+\frac{1}{2} \bar \psi^A \psi^B\epsilon_{AB}\right)\bar\psi ^C\gamma_5\psi^D\epsilon_{CD} \\
        &-\frac{\ii}{2\ell}\bar\psi^A\gamma_{ab}\gamma_5\psi_{A}V^aV^b-\frac{1}{8\ell^2}V^aV^bV^cV^d\epsilon_{abcd} \\ 
        &+\frac{1}{4}\left(\tilde F^{cd}V^aV^b F-\frac{1}{12}\tilde F_{lm} \tilde F^{lm}V^aV^bV^cV^d\right)\epsilon_{abcd} \\
        & -\frac{\ell^2}{8}\left(\mathcal R^{ab} \mathcal R^{cd}\epsilon_{abcd}+\frac{8{\ii}}{\ell} \bar{\rho}^A\gamma_5 \rho_{A}-\frac{2 \ii}{\ell}\mathcal R^{ab}\bar\psi^A\gamma_{ab}\gamma_5\psi_{A}+\frac{4 \ii}{\ell^2}\diff A \bar\psi^A\gamma_5\psi^B\epsilon_{AB}\right)\,.
    \end{split}
\end{equation}
Remarkably, the full Lagrangian \eqref{LfullN2} can be equivalently rewritten in terms of the $\rm{OSp}(2|4)$-covariant supercurvatures, defined as\footnote{As we have done in the $\mathcal{N}=1$ case, we keep on adopting bold symbols to denote the $\rm{OSp}(2|4)$ super field-strengths, without changing, however, their name with respect to the $\mathcal{N}=1$ case, to lighten the notation.}
\begin{equation}\label{lagsuperN2}
\begin{split}
{\bf {R}}^{ab} & \equiv \mathcal{R}^{ab} - \frac{1}{\ell^2} V^a V^b - \frac{1}{2\ell} \delta^{AB} \bar\psi_A \gamma^{ab}\psi_B \,, \\
{\bf {R}}^{a}& \equiv \mathcal{D} V^a -\frac{\ii}{2}\bar\psi^A \gamma^a \psi_A \,, \\
\boldsymbol {{\rho}}_A& \equiv \rho_A  -    \frac{\ii}{2\ell}\delta_{AB}\gamma_a \psi^B V^a \,, \\
\bf{F}& \equiv  F \,.
\end{split}
\end{equation}
In fact, when expressed in terms of the supercurvatures \eqref{lagsuperN2}  the  full Lagrangian acquires the following form:
\begin{equation}\label{MMN2}
    \begin{split}
        \mathcal{L}^{\mathcal{N}=2}_{\text{full}} & = -\frac{\ell^2}{8}{\bf {R}}^{ab}\wedge {\bf {R}}^{cd} \epsilon_{abcd}-\ii\ell\bar{\boldsymbol {{\rho}}}^A \gamma_5 \wedge \boldsymbol {{\rho}}_A+\frac{1}{4} {\bf{F}}\wedge ^*{\bf{F}} \,,
    \end{split}
\end{equation}
where $^*{\bf{F}}$ denotes the Hodge-dual on spacetime of the super field-strength $\bf{F}$.\footnote{The same remark as for the $\mathcal{N}=1$ case, regarding use of the Hodge symbol in superspace, holds here (see footnote \ref{hodge}).}
The Lagrangian \eqref{MMN2} results to be written \`a la MacDowell-Mansouri \cite{MacDowell:1977jt}, that is quadratic in the supercurvatures \eqref{lagsuperN2}, and, in fact, it depends on the fields of the theory only through their $\rm{OSp}(2|4)$-covariant supercurvatures \eqref{lagsuperN2}. Analogously to   the $\mathcal{N}=1$ case, the  resulting MacDowell-Mansouri Lagrangian has an $\rm{OSp}(2|4)$ structure, but it is not $\rm{OSp}(2|4)$ invariant, since its  couplings (in particular the Levi-Civita symbol $\epsilon_{abcd}$ and the matrix $\gamma_5$) are Lorentz invariant tensors, but not $\rm{OSp}(2|4)$ invariant ones.

Finally, as it happened in $\mathcal{N}=1$ for the $\rm{OSp}(1|4)$ supercurvatures, here the field equations on the boundary $\partial \mathcal{M}_4$ can be simply rewritten as the constraint of vanishing at the boundary of the $\rm{OSp}(2|4)$ supercurvatures \eqref{lagsuperN2}, that is
\begin{equation}\label{vanosp24c}
    {\mathbf{R}}^{ab}|_{\partial \mathcal{M}_4} = 0 \,, \quad {\boldsymbol{\rho}}|_{\partial \mathcal{M}_4} = 0 \,, \quad {\bf{F}}|_{\partial \mathcal{M}_4}=0 \,, \quad {\mathbf{R}}^a|_{\partial \mathcal{M}_4}=0 \,.
\end{equation}
Thus, the boundary conditions resulting from the equations of motion, when expressed in terms of four-dimensional superfields and their derivatives, look like Neumann boundary conditions on the supercurvatures \eqref{lagsuperN2}, and are in fact the conditions for global $\mathcal{N}=2$ superconformal invariance of the boundary theory.

We conclude by mentioning that, in the geometric approach to supergravity in superspace, in order to obtain the Lagrangian in ordinary spacetime, one has to perform the restriction $\theta^{\alpha A}=\diff \theta^{\alpha A} = 0$, being $\theta^{\alpha A}$ the fermionic Grassmann coordinates of superspace ($A=1,2$ in the $\mathcal{N}=2$ case, while the index $A$ drops out in $\mathcal{N}=1$), so that the hypersurface, on which we integrate to get the action, immersed in superspace is identified with spacetime (see, e.g., \cite{Castellani:1991et,DAuria:2020guc} and Appendix A of Ref. \cite{Andrianopoli:2014aqa} for details on the derivation of the spacetime Lagrangian from the geometric approach). 

\vskip 5mm
Future developments of the construction we have reviewed in this section may include the extension of the geometric approach to the boundary problem to higher-dimensional, as well as to $\mathcal{N}$-extended, pure or matter coupled, supergravity models including fields with spin lower than one.

In this context, let us stress that the supersymmetric extension of the Euler-Gauss-Bonnet term is unique for a given theory with $\mathcal{N}\geq 2$ supersymmetries, and it is a total derivative, corresponding to a boundary term in superspace. The possible definition of a topological index in superspace associated with this invariant is still an open question. This issue could be properly investigated using the formalism of integral forms in superspace developed in \cite{Castellani:2014goa,Castellani:2015paa}.

\section{Supersymmetry invariance of flat supergravity with boundary}\label{flat}

An interesting question, which was also an open problem of Ref. \cite{Andrianopoli:2014aqa}, is what happens in the vanishing cosmological constant limit ($\Lambda \rightarrow 0$, that is $\ell \rightarrow \infty$) in the presence of a spacetime boundary. In particular, the vanishing cosmological constant limit cannot be naively applied to the supersymmetric (full) MacDowell-Mansouri Lagrangian \eqref{MMN1}, the boundary contributions all vanishing in the limit $\ell \to \infty$.

This issue has been extensively addressed in particular for the case where the boundary is placed asymptotically at infinity. In this case it was shown that  an infinite-dimensional group, named BMS group (and its supersymmetric extensions), emerges as asymptotic symmetry \cite{Bondi:1962px,Sachs:1962,Sachs:1962zza,Barnich:2009se,Barnich:2010eb,Barnich:2011mi,Barnich:2013axa,Strominger:2013jfa,Duval:2014uva,Barnich:2016lyg,Bagchi:2016bcd,Bagchi:2019clu,Aneesh:2021uzk,Gupta:2021cwo}.

Concerning $D=4$ supergravity with boundary, whose construction in the geometric approach has been   reviewed in Sections \ref{geombdy} and \ref{aads4},  a question that naturally arises  is whether, in the limit case of vanishing cosmological constant,  a geometric boundary Lagrangian $\mathcal{L}_{\text{bdy}}$ exhibiting super-BMS symmetry exists.
To answer this question one should first of all   consider  the case of a boundary placed at  asymptotic infinity, in such a way to allow the BMS symmetry to possibly emerge.  On the other hand, in order to implement the geometric approach scheme to the boundary theory, the kind and position of the boundary is not required to be specified, all the results being expressed in terms of tensors with $\rm{SO}(1,3)$ covariant indices.

\vskip 5mm

In the following, we are going to review how the boundary contributions can be taken into account in the geometric approach, disregarding here an explicit choice of the boundary. This will be the subject of Section \ref{flatsugra}.

However, the choice of a boundary is crucial in applications of the formalism to a holographic context. 
Before proceeding with the geometric analysis of pure $\mathcal{N}=1$, $D=4$ ``flat supergravity''\footnote{Recall that here with the terminology ``flat supergravity'' we mean
supergravity in the absence of any explicit internal scale in the Lagrangian.}, let us first briefly sketch, in the following Section \ref{asympt}, some key aspect of the asymptotic symmetries of asymptotically flat spacetimes.

\subsection{Symmetry structure of asymptotically flat spacetimes}\label{asympt}

Asymptotic symmetries of asymptotically flat spacetimes have been subject of great interest in recent years (the literature on this topic is huge; let us refer the reader to, e.g., \cite{Bondi:1962px,Sachs:1962,Sachs:1962zza,Barnich:2009se,Barnich:2010eb,Barnich:2011mi,Barnich:2013axa,Strominger:2013jfa,Duval:2014uva,Barnich:2016lyg,Bagchi:2016bcd,Bagchi:2019clu,Aneesh:2021uzk,Gupta:2021cwo}).
In particular, they play an important role in the formulation of a holographic description of gravity in this case. 
We will further elaborate on this issue  in Section \ref{applicationssect}.

It is still unclear which is the \textit{maximal} set of symmetries admitted by a four-dimensional theory including gravity with asymptotically locally flat boundary conditions.
On the other hand, it is known that with Dirichlet-type boundary conditions, where the non-degenerate spatial part of the boundary metric is taken to be a round 2-sphere, the asymptotic symmetry algebra is the so-called $\mathfrak{bms}_4$ algebra, which includes the Poincaré algebra as its maximal finite-dimensional subalgebra.

In a theory of gravity, the \textit{asymptotic symmetry group} describes the symmetries at the boundary of spacetime. Often, it is larger than the isometry group of the vacuum state of the bulk.\footnote{We refer the reader to \cite{Aneesh:2021uzk} for a meticulous definition of asymptotic symmetry group, quotient of the group of residual gauge transformations (or diffeomorphisms, for theories of gravity) modulo the group of trivial gauge transformations (or, again, diffeomorphisms, for theories of gravity). In particular, one should more properly state that the asymptotic symmetry group of a class of theories is the union of the asymptotic symmetry groups of each (equivalent) formulation of that theory, enforcing in this way the actual asymptotic symmetry group to be invariant under field redefinitions and gauge choices.} For instance, in AdS$_3$ the asymptotic symmetry is enhanced to two copies of the infinite-dimensional Virasoro algebra \cite{Brown:1986nw}, and this infinite-dimensional enhancement of symmetries is at the core of the well-celebrated AdS$_3$/CFT$_2$ correspondence. Another remarkable example is the case of asymptotically flat spacetimes at null (i.e., light-like) infinity. 
There, the asymptotic symmetry group is the infinite-dimensional Bondi-Metzner-Sachs (BMS) group \cite{Bondi:1962px,Sachs:1962,Sachs:1962zza}, instead of the conventionally expected Poincaré group. In particular, the asymptotic symmetry of four-dimensional flat spacetime is referred to as the BMS$_4$ symmetry group. 

The BMS group consists of the semi-direct product of the group of globally defined conformal transformations of the unit 2-sphere (isomorphic to the orthochronous homogeneous Lorentz group) times the infinite-dimensional Abelian normal subgroup of so-called \textit{supertranslations}. The latter are translations along the null direction which depend on the angles of the sphere at infinity. In particular, the Abelian group of supertranslations is infinite-dimensional so that the topology that makes BMS a continuous group is not unique. In physical terms, the supertranslations arise because there are infinitely many directions from which observers at infinity, which are not synchronized and whose world lines coincide with the null generators of null infinity in a certain limit, can observe the system and because each observer is free to choose its own origin of proper time \cite{Frauendiener:2006}. A supertranslation is a shift of the parameter along each null generator of  null infinity corresponding to a change of origin for each individual observer. A choice of origin on each null generator of null infinity is referred to as a ``cut'' of the latter: it is a two-dimensional surface of spherical topology which intersects each null generator exactly once.\footnote{Further enhancements of the asymptotic symmetries of flat spacetime, extending the BMS group by including \textit{superrotations} and \textit{superboosts}, were considered in \cite{Barnich:2009se,Barnich:2010eb} and \cite{Flanagan:2015pxa}.} 

The BMS group is expected be a symmetry of the \textit{gravitational S-matrix} \cite{Strominger:2013jfa}.
In particular, the proposal in \cite{Strominger:2013jfa} is that this is the case for an infinite-dimensional subgroup of the full BMS group (a certain combination, also referred to as the \textit{diagonal BMS group}, of the group acting at future null infinity and the one acting at past null infinity).\footnote{In Strominger's work \cite{Strominger:2013jfa} the BMS transformations acting on future null infinity ($\mathcal{I}^+$) are denoted by BMS$^+$. They are an infinite-dimensional set of ``large'' diffeomorphisms transforming one asymptotically flat solution of general relativity constraints at $\mathcal{I}^+$ to a new, physically inequivalent solution. There is then an isomorphic structure at past null infinity ($\mathcal{I}^-$) acted on by a second copy of the group denoted BMS$^-$.} Such diagonal subgroup of the product of the past and future BMS groups results therefore to be a symmetry of both classical and quantum gravitational scattering \cite{Strominger:2013jfa}. Intriguing connections among the BMS symmetry group, the S-matrix of a quantum theory of gravity, \textit{soft gravitons}, \textit{memory effects}, and the \textit{black hole information loss paradox} have been put forward in fairly recent literature, and not only in the context of theories of gravity. We refer the reader to, e.g., \cite{Bagchi:2016bcd,Strominger:2017zoo} and references therein for further details on these points.

Besides, the analysis of the asymptotic symmetry in a theory of gravity is extremely useful in the context of holography, in particular concerning the physical aspects of the dual field theory: following the celebrated case of the AdS/CFT duality, from which the idea and tools of holography at the boundary of a gravitational theory arose,  one first of all assumes as a  necessary condition that the asymptotic symmetry group of the bulk dictates the global symmetries of the dual field theory living on the boundary of the spacetime. Therefore, in a holographic formulation of quantum gravity in asymptotically flat spacetimes, one would expect the putative dual field theory to be a BMS invariant theory living on the null boundary of spacetime \cite{Bagchi:2016bcd}. This results to be the guideline for the formulation of the so-called \textit{celestial holography} (see, e.g., \cite{Pasterski:2016qvg,Arkani-Hamed:2020gyp}), which, in short, is the statement that the holographic dual of quantum gravity in asymptotically flat four-dimensional spacetime is a conformal field theory living on the two-dimensional celestial (spatial) sphere.

In all the frameworks sketched above for \textit{flat holography},\footnote{Flat holography   denotes the application of the holographic correspondence to the case of asymptotically flat spacetime.} the asymptotic boundary is a spatial surface placed at null infinity. However, null surfaces have in general a degenerate metric, so that any given holographic model has to deal with a consistent definition of the two-dimensional  induced spatial metric. As pointed out in \cite{Bagchi:2019clu}, one way to deal with this problem consists in considering a systematic singular limit where an infinite boost is implemented  on a space-like surface of a  relativistic field theory. This can be achieved by sending the speed of light in the field theory to zero \cite{LevyLeblond:1965}. At the group theoretical level, this limit corresponds to the contraction of  the Poincaré group to what is known as the Carrollian group.
Fields living on a null hypersurface of spacetime necessarily propagate at the speed of light, and they must therefore be massless. This leads to consider, as underlying symmetry of such a field theory, a conformal extension of the Carroll group. Interestingly, the conformal Carroll group was recently shown to be isomorphic to
the BMS group \cite{Duval:2014uva}. All of this can be seen as a heuristic argument as to why there
is a  BMS symmetry associated to any field theory constructed on a null surface. Such field theories will then be conformal and defined on manifolds where the Lorentzian structure has been replaced by a Carrollian one.

At the supersymmetric level, graded extensions of the $\mathfrak{bms}_{4}$ algebra emerge as asymptotic symmetries of supergravity on asymptotically flat spacetimes. Such extensions are generically named super-$\mathfrak{bms}_{4}$ algebras,  but can contain either a finite number \cite{Awada:1985by,Henneaux:2020ekh} or an infinite number \cite{Avery:2015iix,Fotopoulos:2020bqj,Narayanan:2020amh,Fuentealba:2021xhn} of fermionic generators.  In particular, an $\mathcal{N}=1$ structure with an infinite-dimensional odd sector was shown to appear in \cite{Fotopoulos:2020bqj,Fuentealba:2021xhn,Avery:2015gxa} as asymptotic symmetry algebra of $\mathcal{N}=1$ supergravity at null infinity, a subalgebra of which can also be realized as asymptotic symmetry at spatial infinity \cite{Fuentealba:2021xhn}. There also exists another extension of the $\mathfrak{bms}_{4}$ algebra, inequivalent to the one of \cite{Fotopoulos:2020bqj,Avery:2015gxa}, with only four fermionic generators \cite{Awada:1985by,Henneaux:2020ekh}.

\subsection{Construction of the flat model}\label{flatsugra}

Independently on the location of the boundary, under quite general conditions a  boundary Lagrangian allowing to restore the supersymmetry invariance of pure $\mathcal{N}=1$, $D=4$ ``flat supergravity'' was constructed in \cite{Concha:2018ywv}, within a geometric superspace approach. The theory required the inclusion  of appropriate boundary terms depending on some auxiliary fields: an extra bosonic  gauge field, $A^{ab}_{\mu}=-A^{ba}_{\mu}$, and an extra fermionic one, $\chi_{\mu}$.
The full supersymmetric action, given by the bulk plus boundary contributions, was eventually recast in a MacDowell-Mansouri-like form \cite{MacDowell:1977jt}. 

As we will discuss in the following, the auxiliary 1-form fields $A^{ab}$ and $\chi$ enter the Lagrangian only through boundary contributions (total derivatives), but they  are naturally defined in the whole superspace. Their inclusion in the supergravity multiplet allows the matching off-shell of the bosonic and fermionic degrees of freedom in the bulk.
From the bulk perspective, they have the role of implementing the consistency of the  theory, since their field equations are the Bianchi identities associated with the spin connection and gravitino 1-forms. On the other hand,  as proven in \cite{Concha:2018ywv}, 
the condition for the theory to be supersymmetry invariant in the bulk plus boundary  enforces the auxiliary fields to be, on the boundary, the Maurer-Cartan 1-forms  of   a rigid super-Maxwell algebra.

We will also review, following \cite{Concha:2018ywv},  how the   structure sketched above can be obtained in the $\ell \rightarrow \infty$ limit of a deformation of AdS$_4$ supergravity involving  fields redefinition and exhibiting a generalized cosmological constant. A thorough understanding of the physical meaning of the extra fields $A^{ab}$ and $\chi$ under the boundary perspective could be achieved by studying an intrinsic three-dimensional description of the theory, which requires an explicit choice of boundary.
This is beyond the aim of the present review, and will be discussed elsewhere.

Before proceeding with the geometric approach to the boundary problem in the flat supergravity case, let us mention that the inclusion of extra fields in the construction of $\mathcal{N}=1$, $D=4$ flat bulk plus boundary supergravity in the presence of auxiliary fields was also previously considered in \cite{Belyaev:2007bg,Belyaev:2008ex}, but in different terms than what we will see below,
following the tensor calculus approach. More precisely, in \cite{Belyaev:2008ex} a consistent flat supergravity with boundary was constructed within the old-minimal auxiliary fields completion of minimal supergravity. In that model, the extension of the off-shell description with the inclusion of a compensator axial-vector field was required and shown to be associated with  the first component of the extrinsic curvature multiplet.

We are now going to consider flat supergravity on a manifold with boundary and to apply the geometric approach to restore the supersymmetry of the theory by introducing in a geometric way appropriate boundary terms to the Lagrangian. In this way, as shown in \cite{Concha:2018ywv}, the action  including the boundary contributions results to be invariant under supersymmetry transformations.

\vskip 5mm

We focus on the $\mathcal{N}=1$, $D=4$ case, following \cite{Concha:2018ywv}. The field content of the model is the same of Section \ref{N1}. The Lorentz-covariant super field-strengths of the theory are given by \eqref{N1supercurv}.\footnote{Notice that the fermionic bilinear in $R^a$ in \eqref{N1supercurv} has an additional ``$\ii$'' factor with respect to the one of \cite{Concha:2018ywv}. We keep on adopting this ``$\ii$'' factor to be coherent with the conventions adopted in the present paper.}  
The minimal four-dimensional ungauged supergravity Lagrangian, when written as a first-order one, reads\footnote{To be coherent with respect to the notation previously adopted, we multiply the flat supergravity Lagrangian considered in \cite{Concha:2018ywv} by a factor $\frac{1}{4}$.}
\begin{equation}\label{FirstLagr}
\mathcal{L}^{\text{flat}}_{\text{bulk}}= \frac{1}{4} \mathcal{R}^{ab} V^c V^d \epsilon_{abcd} - \bar{\psi} \gamma_5 \gamma_a \rho V^a \,,
\end{equation}
as can also be obtained by taking the flat ($\ell \rightarrow \infty$) limit of \eqref{LbulkN1}.
The Lagrangian \eqref{FirstLagr} is simply given by the Einstein-Hilbert (EH) and Rarita-Schwinger terms, and it scales (in natural units) as $\rm{L}^{2}$, being $\rm{L}^{2}$ the scale-weight of
the EH term. In fact, $[\omega ^{ab}]=\rm{L}^{0}$, $[V^{a}]=\rm{L}$, and $[\psi]=\rm{L}^{1/2}$.

Now, in  the presence of a spacetime boundary, unless trivial boundary conditions are imposed on the fields, the supersymmetry invariance of the theory is broken, and  in order to restore it   we shall add boundary terms to the theory. 
As we have already seen in Section \ref{aads4}, genuine boundary contributions to be considered should scale homogeneously with the other terms of the Lagrangian; in particular they must have the same
scale-weight as the EH term. However, the only Lorentz-invariant boundary terms that can be constructed using the spin-connection $\omega ^{ab}$, the vielbein $V^{a}$, and the gravitino $\psi$ are those given in \eqref{bdyterms1}, which scale with $\rm{L}^{0}$ and $\rm{L}$, respectively in order of appearence in \eqref{bdyterms1}, and the inclusion of such boundary terms leads to the full Lagrangian \eqref{MMN1}, whose direct flat limit does not appear to be well-defined.

The alternative approach proposed in \cite{Concha:2018ywv} consists in adding new gauge fields with higher scale-weight with respect to that of the fields already present in the theory. In particular, the minimal choice to restore supersymmetry invariance is given by the addition of an antisymmetric bosonic 1-form gauge field $A^{ab}=-A^{ba}$ with scale-weight $\rm{L}^2$ and a fermionic 1-form gauge field $\chi$ with scale-weight $\rm{L}^{3/2}$.
The total number of the off-shell d.o.f. of the new gauge fields $A^{ab}_\mu$ is $3\times 6=18$, while that of $\chi_{\mu\alpha}$ is $3\times 4$, exactly as the gravitino. With their inclusion, then, the total number of bosonic bulk d.o.f., including the 6 d.o.f. of the off-shell metric, is $n_B=6+18=24$, matching the total number of fermionic d.o.f. $n_F=12+12=24$. A crucial aspect concerning these auxiliary fields is that their inclusion in the theory allows not only the off-shell matching of the bosonic and fermionic d.o.f. but also to restore the supersymmetry invariance in the context we are considering. In this sense, they play a topological role as they appear only in the boundary Lagrangian necessary to restore supersymmetry in the geometric approach.

Then, the only boundary contributions constructed by using the original field content of the theory together with $A^{ab}$ and $\chi$ that are compatible with parity and Lorentz invariance and that do not involve a scaling parameter are
\begin{equation}\label{bdyt}
    \begin{split}
        & \diff \left( A^{ab}\wedge \mathcal{R}^{cd}+{\omega^a}_f \wedge \omega ^{fb} \wedge A^{cd}+2{\omega^a}_f \wedge A^{fb} \wedge \omega ^{cd}+\omega ^{ab} \wedge \mathcal{F}^{cd}\right) \epsilon_{abcd}=2\mathcal{R}^{ab} \wedge \mathcal{F}^{cd}\epsilon _{abcd}\,, \\
        & \diff \left( \bar{\psi}\gamma _{5}\wedge \sigma +\bar{\chi}\gamma _{5} \wedge \rho \right) =2 \bar{\sigma}\gamma _{5} \wedge \rho -\frac{1}{2}\mathcal{R}^{ab} \wedge \bar{\chi}\gamma_5 \gamma_{ab} \wedge \psi \,,
    \end{split}
\end{equation}
where we have defined
\begin{equation}
    \begin{split}
        \sigma & \equiv \mathcal{D}\chi \,, \\
        \mathcal{F}^{ab} & \equiv \mathcal{D} A^{ab} \,.
    \end{split}
\end{equation}

Thus, the boundary Lagrangian reads
\begin{equation}\label{bdylagr}
\mathcal{L}^{\text{flat}}_{\text{bdy}}= \alpha' \left( 2
\mathcal{R}^{ab}\mathcal{F}^{cd}\epsilon _{abcd}\right) -\ii \beta' \left( 2 \bar{\sigma}\gamma _{5} \rho -\frac{1}{2}\mathcal{R}^{ab} \bar{\chi}\gamma_5 \gamma_{ab} \psi \right) \,,
\end{equation}
where $\alpha'$ and $\beta'$ are constant dimensionless parameters amounting to the normalization of the auxiliary fields, which can be chosen at our wish. Notice that the boundary Lagrangian \eqref{bdylagr} has scale-weight $\rm{L}^{2}$ as the bulk Lagrangian \eqref{FirstLagr}.
Therefore, we are left with the following full Lagrangian:
\begin{equation}\label{full}
\begin{split}
\mathcal{L}^{\text{flat}}_{\text{full}}& =\mathcal{L}^{\text{flat}}_{\text{bulk}}+\mathcal{L}^{\text{flat}}_{\text{bdy}} \\
& = \frac{1}{4} \mathcal{R}^{ab} V^c V^d \epsilon_{abcd} - \bar{\psi} \gamma_5 \gamma_a \rho V^a \\
& + \alpha' \left( 2
\mathcal{R}^{ab}\mathcal{F}^{cd}\epsilon _{abcd}\right) -\ii \beta' \left( 2 \bar{\sigma}\gamma _{5} \rho -\frac{1}{2}\mathcal{R}^{ab} \bar{\chi}\gamma_5 \gamma_{ab} \psi \right) \,.
\end{split}
\end{equation}
Naturally, the boundary terms \eqref{bdyt} do not
affect the bulk and, in particular, we have $\imath _{\epsilon }(\diff \mathcal{L}^{\text{flat}}_{\text{full}})=0$.

Then, the supersymmetry invariance of the full Lagrangian \eqref{full} requires to verify the condition $\imath _{\epsilon }\left( \mathcal{L}^{\text{flat}}_{\text{full}}\right) |_{\partial \mathcal{M}_4}=0$.
As already seen in Section \ref{N1}, also here the field equations acquire non-trivial boundary contributions coming not only from the boundary Lagrangian but also from the bulk one (from the total differentials originating from partial integration), yielding, in particular, the following constraints to hold on the boundary:
\begin{equation}\label{eqbdy}
\begin{cases}
& \mathcal{R}^{ab} |_{\partial \mathcal{M}_4} = 0 \,, \\
& \mathcal{F}^{ab} |_{\partial \mathcal{M}_4} = - \frac{1}{8\alpha'} \left( V^a V^b + \beta' \bar{\chi} \gamma^{ab} \psi \right)_{\partial \mathcal{M}_4} \,, \\
& \rho |_{\partial \mathcal{M}_4} = 0 \,, \\
& \sigma |_{\partial \mathcal{M}_4} = \frac{\ii}{2\beta'} \left( \gamma_a \psi V^a \right)_{\partial \mathcal{M}_4} \,.
\end{cases}
\end{equation}
Thus, the supercurvatures result to be dynamically fixed, on the boundary, to constant values in an enlarged anholonomic basis (indeed, on the boundary they are fixed not only in terms of the supervielbein $\lbrace V^a, \psi \rbrace$ but also of the extra $1$-form field $\chi$).
Computing $\imath _{\epsilon }\left( \mathcal{L}^{\text{flat}}_{\text{full}}\right)$ from \eqref{full} and then considering its projection on the boundary, $\imath _{\epsilon }\left( \mathcal{L}^{\text{flat}}_{\text{full}}\right) |_{\partial \mathcal{M}_4}$, upon use of \eqref{eqbdy} one finds
\begin{equation}
\imath _{\epsilon }\left( \mathcal{L}^{\text{flat}}_{\text{full}}\right) |_{\partial
\mathcal{M}_4}=0 \,.
\end{equation}
Thus, the supersymmetry invariance of the full Lagrangian is restored in the presence of a boundary of spacetime if we include the boundary terms proportional to $\alpha'$ and $\beta'$ (with $\alpha' \neq 0$, $\beta' \neq 0$). The emerging algebraic structure is more transparent once the normalization  coefficients $\alpha'$ and $\beta'$ are fixed to the values $\alpha'=-\frac{1}{8}$ and $\beta'=1$. For these values the full Lagrangian can be rewritten in a MacDowell-Mansouri-like form \cite{MacDowell:1977jt} as\footnote{Notice that actually this is not a MacDowell-Mansouri Lagrangian, as the latter would be quadratic in the field-strengths, while here we have the wedge product of different 2-form curvatures.}
\begin{equation}\label{nMM}
    \mathcal{L}^{\text{flat}}_{\text{full}} = - \frac{1}{4} \mathcal{R}^{ab} \wedge \hat{\mathcal{F}}^{cd}\epsilon _{abcd} - 2 \ii \bar{\Xi} \gamma _{5}\wedge \rho \,,
\end{equation}
where we have defined
\begin{equation}\label{Maxcurv}
    \begin{split}
        \hat{\mathcal{F}}^{ab} &\equiv \mathcal{F}^{ab}-V^{a}V^{b}-\bar{\chi}\gamma ^{ab}\psi \,, \\
        \Xi & \equiv \sigma -\frac{\ii}{2} \gamma_{a} \psi V^a \,.
    \end{split}
\end{equation}
Hence, enlarging the field content of the theory we have been able to restore supersymmetry (which results to be restored for any value of $\alpha'$ and $\beta'$) and, furthermore, to end up with a full Lagrangian in the MacDowell-Mansouri-like form for specific values of $\alpha'$ and $\beta'$.
The enlargement, however, does not modify the bulk Lagrangian, as it affects only the boundary, allowing to restore the supersymmetry invariance.

Remarkably, the supercurvatures \eqref{Maxcurv}, together with (see \eqref{eqbdy}) 
\begin{equation}\label{Maxcurv1}
    \begin{split}
        R^{ab} & \equiv \mathcal{R}^{ab} \,, \\
        \Psi & \equiv \rho \,, \\
        R^a & \equiv \mathcal{D}V^a - \frac{\ii}{2} \bar\psi \gamma^a \psi \,,
    \end{split}
\end{equation}
turn out to reproduce the so-called (minimal) Maxwell-covariant supercurvatures (see, e.g., \cite{Bonanos:2009wy,Concha:2014tca,Concha:2015tla}). The minimal Maxwell superalgebra, dual to the ``vacuum'' Maurer-Cartan structure given by the vanishing of the supercurvatures in  \eqref{Maxcurv} and \eqref{Maxcurv1}, was first introduced in \cite{Bonanos:2009wy} in order
to describe a generalized four-dimensional superspace in the presence of a
constant Abelian supersymmetric field-strength background. Such superalgebra extends the Maxwell algebra, whose generators are $\lbrace J_{ab} , P_a, Z_{ab} \rbrace$, respectively dual to $\omega^{ab}$, $V^a$, and $A^{ab}$, by incorporating two fermionic generators, $Q$ and $\Sigma$, dual to $\psi$ and $\chi$, respectively. At the purely bosonic level, the Maxwell algebra was introduced in \cite{Bacry:1970ye,Schrader:1972zd}. At the supersymmetric level, super-Maxwell algebras were considered in particular in three spacetime dimensions, as they allowed to reproduce three-dimensional Chern-Simons supergravity models in, e.g., \cite{Concha:2015woa,Concha:2018jxx}.\footnote{Further generalizations of the minimal Maxwell superalgebra can be found in \cite{Concha:2014tca,deAzcarraga:2012zv,deAzcarraga:2014jpa,Concha:2014xfa,Penafiel:2017wfr,Ravera:2018vra}, together with diverse applications.} The rigid super-Maxwell algebra in four dimensions has the following structure of (anti)commutators:
\begin{equation}
    \begin{split}
        & \left[ J_{ab},J_{cd}\right] \propto \eta _{bc}J_{ad}-\eta _{ac}J_{bd}-\eta_{bd}J_{ac}+\eta _{ad}J_{bc}\,, \\
        & \left[ J_{ab},P_{c}\right] \propto \eta _{bc}P_{a}-\eta _{ac}P_{b}\,, \quad \left[ P_{a},P_{b}\right] \propto Z_{ab}\,, \\
        & \left[ J_{ab},Z_{cd}\right] \propto \eta _{bc}Z_{ad}-\eta _{ac}Z_{bd}-\eta_{bd}Z_{ac}+\eta _{ad}Z_{bc}\,, \\
        & \left[ J_{ab},Q\right] \propto \gamma_{ab} Q\,, \quad \left[ J_{ab},\Sigma \right] \propto \gamma_{ab} \Sigma \,, \quad \left[ P_{a},Q\right] \propto \gamma_a \Sigma \,, \\
        & \left\{ Q , Q \right\} \propto C\gamma ^{a} P_{a}\,, \quad \left\{ Q , \Sigma \right\} \propto C \gamma^{ab} Z_{ab}\,,
    \end{split}
\end{equation}
where $C$ denotes the charge conjugation matrix. 

We can now interpret the boundary constraints  \eqref{eqbdy},
\begin{equation}\label{superMaxwellzerobdy}
    R^{ab}|_{\partial \mathcal{M}_4} = 0\,, \quad \hat{\mathcal{F}}^{ab}|_{\partial \mathcal{M}_4} = 0 \,, \quad \Psi|_{\partial \mathcal{M}_4} = 0 \,, \quad \Xi|_{\partial \mathcal{M}_4} =0 \,,
\end{equation}
as the condition that the super-Maxwell algebra emerges as global symmetry at the boundary. Furthermore, consistency of the bulk theory requires $R^a=0$ and hence, for continuity, we will also require $R^a|_{\partial \mathcal{M}_4} = 0$, as we will see below.

Then, the full Lagrangian \eqref{nMM} can be rewritten in terms of the Maxwell supercurvatures given in \eqref{Maxcurv} and \eqref{Maxcurv1} as
\begin{equation}\label{flatfullLagr}
\mathcal{L}^{\text{flat}}_{\text{full}}= {-} \frac{1}{4} R^{ab} \wedge \hat{\mathcal{F}}^{cd}\epsilon_{abcd} - 2 \ii \bar{\Xi} \gamma _{5} \wedge \Psi\,.
\end{equation}
Note that $A^{ab}$ and $\chi$ appear only in the boundary Lagrangian, but they act as auxiliary fields under the bulk perspective, implementing the Bianchi identities of Lorentz and supersymmetry respectively, associated with $\omega^{ab}$ and $\psi$. Indeed, their equations of motion (e.o.m.) yield, respectively (up to boundary terms),
\begin{equation}
    \begin{split}
        \text{e.o.m. } A^{ab} \quad & \leftrightarrow \quad \mathcal{D} \mathcal{R}^{ab} = 0 \,, \\
        \text{e.o.m. } \chi \quad & \leftrightarrow \quad \mathcal{D} \rho - \frac{1}{4} \mathcal{R}^{ab} \gamma_{ab} \psi = 0 \,.
    \end{split}
\end{equation}
On the other hand, those of $\omega^{ab}$ and $\psi$ read
\begin{equation}\label{eom2}
    \begin{split}
        \text{e.o.m. } \omega^{ab} \quad & \leftrightarrow \quad \mathcal{D} \hat{\mathcal{F}}^{ab} - 2 {R^{[a}}_c A^{c|b]} + \bar{\Xi} \gamma^{ab} \psi - \bar{\chi} \gamma^{ab} \Psi = 0 \,, \\
        \text{e.o.m. } \psi \quad & \leftrightarrow \quad \mathcal{D} \Xi - \frac{1}{4} R^{ab} \gamma_{ab} \chi + \frac{\ii}{2} \gamma_a \Psi V^a = 0 \,.
    \end{split}
\end{equation}
in a kind of, let us say, ``symmetric way'' with respect to the e.o.m. of $A^{ab}$ and $\chi$.

Comparing the latter with the Bianchi identities of the super-Maxwell algebra associated with the auxiliary fields, we can see that, imposing the supertorsion constraint $R^a=0$, they coincide. In this sense, the supertorsion constraint appears as a consistency requirement for the full MacDowell-Mansouri theory (while, as usual, it can be proven that it naturally emerges from the study of the field equations of the bulk Lagrangian in the absence of the boundary contribution $\mathcal{L}^{\text{flat}}_{\text{bdy}}$). Hence, the fields equations of $\omega^{ab}$ and $\psi$ implement the Bianchi identities of the super-Maxwell algebra associated with $A^{ab}$ and $\chi$, respectively. Let us observe that 
the Rarita-Schwinger  equation of motion of the gravitino is hidden in the second of the above equations.  It can be retrieved if we restrict the auxiliary field $\chi$ to be defined  only on the boundary, in which case,  in the bulk,  $\chi= 0$, and the second equation in \eqref{eom2} reduces, in the bulk, to the  Rarita-Schwinger equation.

Finally, the field equations of $V^a$  read
\begin{equation}
    \frac{1}{2}V^b \mathcal{R}^{cd} \epsilon_{abcd}-\bar{\psi}\gamma_a \gamma_5 \rho=0 \,.
\end{equation}
They are the Einstein equations  in superspace, written in the Einstein-Cartan formalism, with energy momentum tensor associated with the  propagating gravitino field.

We observe that the full Lagrangian \eqref{nMM} cannot be directly obtained as a flat limit of \eqref{MMN1}. In particular, in the present case we have new 1-form gauge fields with the associated field-strengths contributing to the full Lagrangian. Nevertheless, as we are going to review in the following Section \ref{AdSLandflat}, the Lagrangian \eqref{nMM}  can be retrieved as zero cosmological constant  (that is infinite AdS radius)  limit  $\ell \rightarrow \infty$ from a theory originating from AdS$_4$ supergravity, but with super AdS-Lorentz covariance, where the extra 1-form gauge fields do not appear only in the boundary Lagrangian but also in the bulk one.

\subsection{Recovering flat supergravity with boundary from super-AdS$_4$}\label{AdSLandflat}

Coupling of the gravitational field with bosonic 1-form fields carrying Lorentz indices typically arises in generalized models of (super)gravity,  where an internal scale  parameter, $\ell$, can be introduced and associated with a generalized definition of  cosmological constant (see, e.g., \cite{Ipinza:2016con,Banaudi:2018zmh,Concha:2015tla,deAzcarraga:2010sw,Penafiel:2018vpe}). The structure group in this case is given by a super AdS-Lorentz group, which can be found  by performing a particular algebraic expansion, named S-expansion \cite{Izaurieta:2006zz}, of the $\mathcal{N}$-extended $\rm{OSp}(\mathcal{N},4)$ super AdS$_4$ group.

One can then construct a Lagrangian covariant under the super AdS-Lorentz group in the presence of a spacetime boundary. Remarkably, the $\ell \rightarrow \infty$ limit of the above theory is able to reproduce the full Lagrangian \eqref{flatfullLagr} of flat supergravity, showing also  how the auxiliary fields $A^{ab}$ and $\chi$  emerge in this limit, out of the bulk fields of the  theory with generalized cosmological constant.

Let us discuss here how the generalized supergravity model discussed in the present section arises from standard supergravity, and why the inherent deformation of the Lorentz structure group is particularly useful in order to reproduce, in the zero cosmological constant limit, the theory of flat supergravity with boundary discussed in Section \ref{flatsugra}.
 
As already emphasized, the full Lagrangian of flat supergravity with boundary cannot be found as a straightforward limit from the one of pure AdS$_4$ supergravity, the limit being singular and the boundary contributions all being vanishing in the limit. Nevertheless, as we are going to discuss in the following, in \cite{Concha:2018ywv} a well-defined zero cosmological constant limit from pure AdS$_4$ supergravity was engineered, by allowing different scalings of the various fields of the theory and using the gained flexibility to write a boundary Lagrangian with a well-defined  limit $\ell \to \infty$.
 
To clarify the procedure, let us start from the AdS$_4$ supergravity theory discussed in Section \ref{N1}, with supecurvatures defined as in \eqref{N1supercurv}, and perform the following field redefinitions, both vanishing in the zero cosmological constant limit $\ell \to \infty$:
\begin{itemize}
\item We introduce a torsionful spin connection 
\begin{equation}
     \hat\omega^{ab}\equiv \omega^{ab} + \frac{1}{\ell^2} A^{ab},
\end{equation}
where $A^{ab}$ is a tensor 1-form, so that
\begin{equation}\label{torful}
     \begin{split}
     \mathcal{R}^{ab}& \to \hat{\mathcal{R}}^{ab} = \diff \omega^{ab}+{\omega^a}_c \omega^{cb}+\frac{1}{\ell^2} \mathcal{D}_{(\omega)} A^{ab}+\frac{1}{\ell^{4}} {A^a}_c A^{cb} \equiv \mathcal{R}^{ab}+\frac{1}{\ell^2} \mathbb{F}^{ab} \,, \\
     {R}^{a}& \to \hat{R}^{a} =  \mathcal{D}_{(\omega)} V^a  +\frac{1}{\ell^2}{A^a}_{b} V^{b}- \frac{\ii}{2} \bar\psi \gamma^a \psi \,,
     \end{split}
\end{equation}
where 
$$\mathbb{F}^{ab} \equiv \mathcal{D}_{(\omega)}A^{ab}+\frac{1}{\ell^2}{A^a}_c A^{cb}\,.$$
\item We redefine the gravitino 1-form with the introduction of the new spinor 1-form $\chi$,
\begin{equation}
    \psi \to \psi +\frac 1\ell \chi \,,
\end{equation}
so that
\begin{equation}\label{deftor}
    \begin{split}
      \hat R^a &\to \mathfrak{R}^{a} \equiv \mathcal{D}_{(\omega)} V^a - \frac{\ii}{2} \bar\psi \gamma^a \psi +\frac{1}{\ell^2}{A^a}_{b} V^{b}-\frac{\ii}{\ell}\bar{\psi}\gamma ^{a}\chi -\frac{\ii}{2\ell ^{2}}\bar{\chi}\gamma ^{a}\chi \,, \\
        \rho &\to \hat\rho = \mathcal{D}_{(\omega)}\psi + \frac 1\ell\left(
        \mathcal{D}_{(\omega)}\chi +\frac{1}{4\ell }A^{ab}\gamma _{ab}\psi +\frac{1}{4\ell ^{2}} A^{ab}\gamma _{ab}\chi \right)\equiv \rho + \frac 1\ell\Phi \,,
    \end{split}
\end{equation}
where
$$\Phi\equiv  \mathcal{D}_{(\omega)}\chi +\frac{1}{4\ell }A^{ab}\gamma _{ab}\psi +\frac{1}{4\ell ^{2}} A^{ab}\gamma _{ab}\chi\,.$$
\end{itemize}
All together, the redefined super field-strengths read
\begin{equation}\label{Ltc}
    \begin{split}
        \mathcal{R}^{ab}& \equiv \diff \omega ^{ab}+{\omega^a}_c \omega ^{cb} \,, \\
        \mathfrak{R}^{a}& \equiv \mathcal{D} V^a - \frac{\ii}{2} \bar\psi \gamma^a \psi +\frac{1}{\ell^{2}}{A^a}_{b} V^{b}-\frac{\ii}{\ell}\bar{\psi}\gamma ^{a}\chi -\frac{\ii}{2\ell ^{2}}\bar{\chi}\gamma ^{a}\chi \,, \\
       \rho & \equiv \mathcal{D}\psi \,, \\
        \mathbb{F}^{ab}& \equiv \mathcal{D}A^{ab}+\frac{1}{\ell ^{2}}{A^a}_c A^{cb}\,, \\
        \Phi & \equiv \mathcal{D}\chi +\frac{1}{4\ell }A^{ab}\gamma _{ab}\psi +\frac{1}{4\ell ^{2}} A^{ab}\gamma _{ab}\chi \,. 
    \end{split}
\end{equation}
In terms of the redefined fields and of their field-strengths, the  bulk Lagrangian \eqref{LbulkN1} of pure AdS$_4$ supergravity was constructed in \cite{Concha:2018ywv}\footnote{Here again we adapt the conventions in such a way to be consistent with the ones previously adopted in the present paper.} and it
reads 
\begin{equation}\label{bulkL}
\begin{split}
\mathcal{L}^{\ell}_{\text{bulk}} & = \frac{1}{4} \epsilon _{abcd}R^{ab}V^{c}V^{d}+\frac{1}{4\ell ^{2}} \epsilon _{abcd}\mathbb{F}^{ab}V^{c}V^{d}-\bar{\psi}\gamma
_{5}\gamma _{a}\rho V^{a}-\frac{1}{\ell} \bar{\psi}\gamma _{5}\gamma _{a}\Phi V^{a} \\
&-\frac{1}{\ell ^{2}} \bar{\chi}\gamma _{5}\gamma _{a}\Phi V^{a}-\frac{1}{\ell} \bar{\chi}\gamma _{5}\gamma _{a}\rho V^{a}-\frac{1}{8\ell ^{2}}\epsilon _{abcd}V^{a}V^{b}V^{c}V^{d} \\
&-\frac{\ii}{2\ell} \bar{\psi} \gamma_5\gamma _{ab}\psi V^{a}V^{b}-\frac{\ii}{\ell ^{2}} \bar{\chi}\gamma_5\gamma _{ab}\psi V^{a}V^{b}-\frac{\ii}{2\ell ^{3}} \bar{\chi}\gamma_5\gamma _{ab}\chi V^{a}V^{b}\,.
\end{split}
\end{equation}

Thinking now at $\psi$ and $\chi$ as independent odd directions of an enlarged superspace, the Lagrangian \eqref{LbulkN1} is left invariant under generalized supersymmetry transformations $\epsilon, \varepsilon$, associated with diffeomorphisms in the directions of $\psi$ and $\chi$, respectively. They are given by
\begin{eqnarray}
\left\{\begin{array}{lll}
\delta_\epsilon \omega^{ab}&=& \ii \left[ \bar\rho^{ab} \gamma_c-2\bar\rho^{[a}_{\phantom{[a}c}\gamma^{b]} + \frac{1}{\ell} \left( \bar\Phi^{ab} \gamma_c-2\bar\Phi^{[a}_{\phantom{[a}c}\gamma^{b]} \right) \right] \epsilon V^c + \frac 1{\ell} \bar \epsilon \gamma^{ab}\psi \,, \\
\delta_\epsilon \psi&=& \mathcal{D}\epsilon \,, \\
\delta_\epsilon V^a &=&\ii \bar \epsilon \gamma^{a}\psi + \frac{\ii}{\ell} \bar{\epsilon} \gamma^a \chi \,, \\
\delta_\epsilon A^{ab} &=& \bar{\epsilon} \gamma^{ab} \chi \,, \\
\delta_\epsilon \chi &=& \frac{\ii}{2} \gamma_a \epsilon V^a + \frac{1}{4\ell} \gamma_{ab} \epsilon A^{ab} \,,
\end{array}\right.
\end{eqnarray}
along with
\begin{eqnarray}
\left\{\begin{array}{lll}
\delta_\varepsilon \omega^{ab}&=& \frac{\ii}{\ell} \left( \bar\rho^{ab} \gamma_c-2\bar\rho^{[a}_{\phantom{[a}c}\gamma^{b]} \right) \varepsilon V^c \,, \\
\delta_\varepsilon \psi&=& 0 \,, \\
\delta_\varepsilon V^a &=& \frac{\ii}{\ell} \bar \varepsilon \gamma^{a}\psi + \frac{\ii}{\ell^2} \bar{\varepsilon} \gamma^a \chi \,, \\
\delta_\varepsilon A^{ab} &=& \ii \left( \bar\Phi^{ab} \gamma_c-2\bar\Phi^{[a}_{\phantom{[a}c}\gamma^{b]} \right) \varepsilon V^c + \bar{\varepsilon} \gamma^{ab} \psi + \frac{1}{\ell} \bar{\varepsilon} \gamma^{ab} \chi \,, \\
\delta_\varepsilon \chi &=& \mathcal{D} \varepsilon + \frac{i}{2\ell} \gamma_a \varepsilon V^a + \frac{1}{4\ell^2} \gamma_{ab} \varepsilon A^{ab} \,,
\end{array}\right.
\end{eqnarray}
where $\rho_{ab}$ and $\Phi_{ab}$ denote the components along the purely bosonic vielbein of the field-strength 2-forms $\rho$ and $\Phi$, respectively, namely the corresponding supercovariant field-strengths. 

The vanishing cosmological constant limit $\ell \rightarrow \infty $ of \eqref{bulkL} yields the flat bulk supergravity Lagrangian \eqref{FirstLagr}, exactly as starting from the equivalent Lagrangian \eqref{LbulkN1}.
However,   the bulk Lagrangian \eqref{bulkL} now includes terms involving the dimensionful 1-form fields $A^{ab}$ and $\chi$.  In the flat supergravity case reviewed in Section
\ref{flatsugra}, the contribution of such 1-forms  to the boundary Lagrangian was crucial to restore supersymmetry invariance in the presence of a boundary.

In \cite{Concha:2018ywv}, the supersymmetry invariance of the theory with generalized cosmological constant was analyzed in the presence of a non-trivial boundary, with the aim of finding a consistently defined flat limit $\ell \rightarrow \infty$ in which the results reviewed in Section \ref{flatsugra} could be recovered. In order to restore supersymmetry, it is necessary to add boundary terms. 
 
In the case at hand, the dimensionful 1-forms $A^{ab}$ and $\chi$ already appear in the bulk Lagrangian \eqref{bulkL}, and they can also be included  within total derivative terms contributing to the boundary Lagrangian, more general to the ones directly obtained from  \eqref{LbdyN1}  when performing the redefinition of spin connection and gravitino. In particular, one can allow  independent couplings in each total derivative term. All possible boundary terms, compatible with the symmetries of the theory, that can be added  (without including extra new fields) yield the following boundary Lagrangian:
\begin{equation}\label{bdyLagrangian}
    \begin{split}
        \mathcal{L}^{\ell}_{\text{bdy}} & = \lambda \epsilon _{abcd}R^{ab}R^{cd} - \ii \pi \left(
        \bar{\rho}\gamma _{5}\rho -\frac{1}{4}R^{ab}\bar{\psi}\gamma_5 \gamma_{ab} \psi
         \right) +\mu \epsilon_{abcd} \left(2 R^{ab} \mathbb{F}^{cd} + \frac{1}{\ell^{2}}\mathbb{F}^{ab}\mathbb{F}^{cd}\right) \\
        & - \ii \nu \left( 2\bar{\rho}\gamma _{5}\Phi  +\bar{\Phi}\gamma _{5}\Phi -\frac{1}{2}R^{ab}\bar{\psi}\gamma _5 \gamma_{ab} \chi - \frac{1}{4\ell}\mathbb{F}^{ab}\bar{\psi}\gamma_5 \gamma_{ab}\psi \right. \\
        & \left. -\frac{1}{2\ell^{2}}\mathbb{F}^{ab}\bar{\psi}\gamma_5\gamma _{ab}\chi -\frac{1}{4\ell}R^{ab}\bar{\chi}\gamma_5 \gamma _{ab}\chi - \frac{1}{4\ell^{3}}\mathbb{F}^{ab}\bar{\chi}\gamma_5 \gamma _{ab}\chi \right) \,, 
    \end{split}
\end{equation}
where $\lambda$, $\pi$, $\mu$, $\nu$ are independent constant parameters. Let us remark that a necessary condition, in order to have a consistently defined limit $\ell \rightarrow \infty $ at the level of the full Lagrangian, is that the Lagrangian should not contain terms diverging in the limit.
For this reason, 
terms involving positive powers of $\ell $ must be excluded. Inspecting \eqref{bdyLagrangian}, all terms  scale as a length squared, $\rm{L}^{2}$, with the exception of those proportional to $\lambda$ and $\pi$. In order to have an appropriately scaled boundary Lagrangian one should define new dimensionless constants, $\lambda^{\prime}$ and $\pi^{\prime}$, such that $\lambda =\ell
^{2}\lambda ^{\prime }$ and $\pi =\ell \pi ^{\prime }$. However, this implies positive powers of $\ell$ in \eqref{bdyLagrangian}. Thus, we conclude that the contributions proportional to $\lambda$ and $\pi$ must be dropped out from the boundary Lagrangian. Finally,  we are left with the full Lagrangian 
\begin{equation}\label{FullL}
    \begin{split}
        \mathcal{L}^{\ell}_{\text{full}} & = \mathcal{L}^{\ell}_{\text{bulk}} + \mathcal{L}^{\ell}_{\text{bdy}} \\
        & = \frac{1}{4} \epsilon _{abcd}R^{ab}V^{c}V^{d}+\frac{1}{4\ell ^{2}} \epsilon_{abcd}\mathbb{F}^{ab}V^{c}V^{d}-\bar{\psi}\gamma _{5}\gamma _{a}\rho V^{a}-\frac{1}{\ell} \bar{\psi}\gamma _{5}\gamma _{a}\Phi V^{a} \\
        &-\frac{1}{\ell ^{2}} \bar{\chi}\gamma _{5}\gamma _{a}\Phi V^{a}-\frac{1}{\ell} \bar{\chi}\gamma _{5}\gamma _{a}\rho V^{a}-\frac{1}{8\ell ^{2}}\epsilon _{abcd}V^{a}V^{b}V^{c}V^{d} \\
        &-\frac{\ii}{2\ell} \bar{\psi} \gamma_5\gamma _{ab}\psi V^{a}V^{b}-\frac{\ii}{\ell ^{2}} \bar{\chi}\gamma_5\gamma _{ab}\psi V^{a}V^{b}-\frac{\ii}{2\ell ^{3}} \bar{\chi}\gamma_5\gamma _{ab}\chi V^{a}V^{b} \\
        & +\mu \epsilon_{abcd} \left(2 R^{ab} \mathbb{F}^{cd} + \frac{1}{\ell^{2}}\mathbb{F}^{ab}\mathbb{F}^{cd}\right) \\
        & - \ii \nu \left( 2\bar{\rho}\gamma _{5}\Phi  +\bar{\Phi}\gamma _{5}\Phi -\frac{1}{2}R^{ab}\bar{\psi}\gamma _5 \gamma_{ab} \chi - \frac{1}{4\ell}\mathbb{F}^{ab}\bar{\psi}\gamma_5 \gamma_{ab}\psi \right. \\
        & \left. -\frac{1}{2\ell^{2}}\mathbb{F}^{ab}\bar{\psi}\gamma_5\gamma _{ab}\chi -\frac{1}{4\ell}R^{ab}\bar{\chi}\gamma_5 \gamma _{ab}\chi - \frac{1}{4\ell^{3}}\mathbb{F}^{ab}\bar{\chi}\gamma_5 \gamma _{ab}\chi \right) \,.
    \end{split}
\end{equation}
Again, from the study of the field equations in the presence of a boundary, the supercurvatures \eqref{Ltc} result to be fixed, on the boundary, to constant values in an enlarged anholonomic basis given by the supervielbein together with the $1$-form field $\chi$,
\begin{equation}\label{curvbdyell}
\begin{cases}
 & R^{ab} |_{\partial \mathcal{M}_4} = - \frac{\nu}{16 \mu \ell} \left( \bar{\psi} \gamma^{ab} \psi \right)_{\partial \mathcal{M}_4} \,, \\
 & \mathbf{F}^{ab} |_{\partial \mathcal{M}_4} = - \frac{1}{8\mu} \left( V^a V^b + \nu \bar{\chi} \gamma^{ab} \psi + \frac{\nu}{2\ell} \bar{\chi} \gamma^{ab} \chi \right)_{\partial \mathcal{M}_4} \,, \\
 & \rho |_{\partial \mathcal{M}_4} = 0 \,, \\
 & \Phi |_{\partial \mathcal{M}_4} = \frac{\ii}{2\nu} \left( \gamma_a \psi V^a + \frac{1}{\ell} \gamma_a \chi V^a \right)_{\partial \mathcal{M}_4} \,,
\end{cases}
\end{equation}
and, consequently, the condition $\imath _{\epsilon }\left( \mathcal{L}^{\ell}_{\text{full}}\right) |_{\partial
\mathcal{M}_4}=0$ for supersymmetry of the full Lagrangian is realized when the following relation holds:
\begin{equation}
\nu = - 8 \mu \left( 1+h\right) \,, \quad \text{where} \quad h =\pm \sqrt{ 1+\frac{1}{8\mu }} \in \mathbb{R} \,,
\end{equation}
with $\nu \neq 0$, which implies $h \neq -1$.
We can then solve the above in terms of the real parameter $h$, obtaining
\begin{equation}
\mu =- \frac{1}{8} \frac{1}{1-h^2}\,,\quad \nu =\frac{1}{1-h}\,,   
\end{equation}
which imply
\begin{equation}\label{curvbdyellbis}
\begin{cases}
 & R^{ab} |_{\partial \mathcal{M}_4} =(1+h)  \left[\frac{1}{2\ell} \bar{\psi} \gamma^{ab} \psi \right]_{\partial \mathcal{M}_4} \,, \\
 & \mathbf{F}^{ab} |_{\partial \mathcal{M}_4} =(1+h)  \left[ (1-h)V^a V^b + \bar{\chi} \gamma^{ab} \psi + \frac{1}{2\ell} \bar{\chi} \gamma^{ab} \chi \right]_{\partial \mathcal{M}_4} \,, \\
 & \rho |_{\partial \mathcal{M}_4} = 0 \,, \\
 & \Phi |_{\partial \mathcal{M}_4} = (1-h)\frac{{\ii}}{2} \left[ \gamma_a \psi V^a + \frac{1}{\ell} \gamma_a \chi V^a \right]_{\partial \mathcal{M}_4} \,.
\end{cases}
\end{equation}

Observe that, setting $h=0$, we obtain
\begin{equation}\label{adsLvalues}
\mu =- \frac{1}{8}\quad \Rightarrow \quad \nu = 1 \,.
\end{equation}
Remarkably, with these values of $\mu$ and $\nu$, the full Lagrangian \eqref{FullL} acquires a MacDowell-Mansouri-like form, that is
\begin{equation}\label{finalFULLLagr}
\mathcal{L}^{\ell}_{\text{full}}= - \frac{1}{4} \mathfrak{R}^{ab}\mathfrak{F}^{cd}\epsilon _{abcd}-
\frac{1}{8\ell ^{2}}\mathfrak{F}^{ab}\mathfrak{F}^{cd}\epsilon _{abcd}-2 \ii \bar{\Omega}\gamma _{5}\rho -\frac{\ii}{\ell }\bar{\Omega}\gamma _{5}\Omega \,,  
\end{equation}
written in terms of the following super field-strengths:
\begin{equation}\label{sc1}
    \begin{split}
        \mathfrak{R}^{ab}& \equiv \diff \omega^{ab} + {\omega^a}_c \omega^{cb} - \frac{1}{2\ell}\bar{\psi}\gamma ^{ab}\psi \,, \\
        \mathfrak{R}^{a}& \equiv \mathcal{D} V^a - \frac{\ii}{2} \bar\psi \gamma^a \psi +\frac{1}{\ell^{2}}{A^a}_{b} V^{b}-\frac{\ii}{\ell}\bar{\psi}\gamma ^{a}\chi -\frac{\ii}{2\ell ^{2}}\bar{\chi}\gamma ^{a}\chi \,, \\
        \rho & \equiv \mathcal{D} \psi \,, \\
        \mathfrak{F}^{ab}& \equiv \mathcal{D}A^{ab} - V^a V^b - \bar\chi \gamma^{ab} \psi +
        \frac{1}{\ell ^{2}}{A^a}_{c} A^{cb}-\frac{1}{2\ell}\bar{\chi}\gamma^{ab}\chi \,, \\
        \Omega & \equiv \mathcal{D}\chi - \frac{\ii}{2} \gamma_a \psi V^a - \frac{\ii}{2\ell} \gamma_a \chi V^a + \frac{1}{4\ell} A^{ab} \gamma_{ab} \psi + \frac{1}{4\ell^{2}}A^{ab}\gamma _{ab}\chi \,.
    \end{split}
\end{equation}
In particular, considering the case $h=0$, from the boundary conditions \eqref{curvbdyellbis}, arising from the study of the field equations in the presence of a boundary, one can prove that the supercurvatures defined in \eqref{sc1} vanish at the boundary. 
Thus we have recovered the supersymmetry invariance of the theory with generalized cosmological constant in the presence of a non-trivial boundary of spacetime and, furthermore, the flat limit $\ell \rightarrow \infty $ is now well-defined in the MacDowell-Mansouri-like formalism. Indeed, the vanishing cosmological constant limit of the full Lagrangian \eqref{finalFULLLagr} yields precisely the flat supergravity model with boundary that we derived in Section \ref{flatsugra}. In particular, in this case not only the bulk Lagrangians are properly related through the flat limit but also the boundary contributions. Moreover, the supercurvatures \eqref{sc1} boil down to the super-Maxwell curvatures \eqref{Maxcurv} and \eqref{Maxcurv1} in the flat limit $\ell \rightarrow \infty$. 

Let us conclude this analysis by observing that setting the right-hand side of supercurvatures \eqref{sc1} to zero, which defines the vacuum of the theory, we obtain the Maurer-Cartan equations associated with a supersymmetric extension of the so-called AdS-Lorentz algebra. Such bosonic algebra is a semi-simple extension of the Poincaré algebra and it was first introduced in \cite{Soroka:2006aj,Gomis:2009dm}.\footnote{Generalizations of the AdS-Lorentz algebra have been useful to recover diverse Lovelock theories
from Chern-Simons and Born-Infeld gravity models \cite{Concha:2016kdz,Concha:2016tms,Concha:2017nca}.}
The supersymmetric extension of the AdS-Lorentz algebra obtained in the present case contains two fermionic charges, $Q$ and $\Sigma$, respectively dual to the spinor 1-form fields $\psi$ and $\chi$, as the minimal AdS-Lorentz superalgebras introduced in \cite{Concha:2015tla}. However, it does not involve additional bosonic generators with respect to the ones generating the AdS-Lorentz algebra in order for the (super-)Jacobi identities to be satisfied. 
In this context, let us stress that the supersymmetrization of the AdS-Lorentz algebra is not unique. In fact, for instance, an AdS-Lorentz superalgebra with just one fermionic charge was considered in \cite{Soroka:2006aj,Fierro:2014lka}.
In the limit $\ell \rightarrow \infty $ the AdS-Lorentz superalgebra associated with \eqref{sc1} precisely boils down to the super-Maxwell algebra defining the vacuum structure arising from the vanishing of the super field-strengths \eqref{Maxcurv} and \eqref{Maxcurv1}.

\section{Applications of the formalism to asymptotic boundaries}\label{applicationssect}

The formalism described in the previous sections is very general, and particularly well suited to be adapted to an holographic framework, in the spirit of the AdS/CFT duality \cite{Maldacena:1997re,Gubser:1998bc,Witten:1998qj}.
In all the models considered above, the boundary terms added in the Lagrangian to restore supersymmetry invariance were written in terms of four-dimensional, $\rm{SO}(1,3)$-Lorentz covariant tensor quantities.
However, to properly unveil the role of the various boundary terms, it is necessary to express them as proper three-dimensional contributions,
the result depending on the general features of the boundary considered, and on its space-like, time-like, or light-like nature.

An interesting application was found in \cite{Andrianopoli:2018ymh}, where the results of \cite{Andrianopoli:2014aqa} were applied by choosing a  specific, locally AdS$_3$ boundary geometry: a peculiar three-dimensional theory exhibiting ``unconventional supersymmetry'' \cite{Alvarez:2011gd} (referred to, in the following, as AVZ model, from the names of the authors) was retrieved at the asymptotic boundary of pure $\mathcal{N}=2$ AdS$_4$ supergravity. The AVZ model is based on a $3d$ Chern-Simons (CS) Lagrangian with $\rm{OSp}(2|2)$ supergroup, but features a  Dirac spinor, that we are going to call $\chi^{\text{(AVZ)}}$,\footnote{Note that the field $\chi^{\text{(AVZ)}}$ is a Dirac spinor, not to be confused with the 1-form field $\chi$ of Section \ref{flat}.} as the only propagating degree of freedom. The AVZ model is interesting, among other peculiar features, because it was proven to have important applications in
condensed matter physics, in particular in the description of graphene-like systems near the Dirac points. The Dirac spinor $\chi^{\text{(AVZ)}}$ of the AVZ model emerges  by imposing  the condition on the spacetime component of the odd CS connection 1-form $\Psi$
\begin{equation}\label{avz}
\chi^{\text{(AVZ)}}_{\alpha} =\ii \left(\gamma^i\right)_{\alpha \beta}\Psi^\beta_{\mu} e^\mu_i\,,
\end{equation}
$e^\mu_i$ being the (inverse) dreibein of the base spacetime where the CS action is integrated on, and $\gamma^i$ a set of gamma matrices on it (while, here, $\alpha,\beta,\ldots=1,2$ are spinorial $\rm{SL}(2,\mathbb{R})$ Lorentz indices, $i=0,1,2$ anhonolonomic $\rm{SL}(2,\mathbb{R})$ vector indices, and $\mu=0,1,2$ the corresponding curved indices). The above ansatz can be read as the condition that the spin-$1/2$ part (associated with gauge freedom), out of the spin-$(1\times 1/2)=3/2 +1/2$ component on the base space, $\Psi_{\mu}^\beta$, of the odd generator, is propagating on the base space. The same relation also requires to choose a metric on the base space. Note that, in $3d$ Chern-Simons theories, the two features mentioned above (introduction of a background dependence, and local dynamics) are both typical of a (even partial) gauge fixing of the gauge symmetry of the CS connection. Indeed, as pointed out in \cite{Andrianopoli:2019sqe}, the relation \eqref{avz} can be interpreted in terms of a particular gauge fixing of the odd symmetry in the CS action. Such condition, however, identifies indices in the spinor representation of the gauge supergroup, carried by the odd generator $\Psi^\beta$, with spinor indices on the  base space, that is it identifies the subgroup $\rm{SL}(2,\mathbb{R})\subset \rm{OSp}(2|2)$ of the gauge supergroup with the Lorentz group $\rm{SL}(2,\mathbb{R})_L$ on the base space.
It therefore defines a non-trivial sector of the classical CS theory. 
The correspondence was then extended in \cite{Andrianopoli:2019sip} to the case $\mathcal{N}>2$, assuming trivial boundary conditions to hold for supergravity fields of spin lower than $1$. This allowed, as a first result, to give a precise map of the Semenoff
and Haldane-type masses appearing in the Dirac equation satisfied by the massive Dirac spinor of the AVZ model extended to $\mathrm{OSp}\left(\frac{\mathcal{N}}2|2\right)\times \mathrm{OSp}\left(\frac{\mathcal{N}}2|2\right)$ in terms of the torsion parameters of the model, on the lines of \cite{Hughes:2012vg,Parrikar:2014usa}. 
To have retrieved the defining equations of the AVZ model, with its intriguing condensed matter implications, at the boundary of AdS$_4$ supergravity is an interesting achievement in itself. However, it is still to be understood whether it is possible to establish a precise holographic duality relation between AdS$_4$ supergravity at large radius and the AVZ model at the quantum level, in the spirit of the AdS/CFT duality. In particular, in this case it is not clear whether the quantum boundary theory can still be associated with a conformal field theory, since the mass term of the Dirac spinor, needed for the correspondence with the gravity theory to take place, would plausibly break the conformal invariance.

Independently from the possibility of a holographic description of the AVZ model, the $\mathcal{N}=2$ model presented in Section \ref{aads4} is particularly well suited to setup a holographic investigation, by applying the holographic renormalization scheme \cite{Skenderis:2002wp}, the condition for regularizing the four-dimensional action being in fact, as we have seen, the request of superconformal invariance of the boundary theory. As we have already discussed in the preamble of the present review, at the bosonic level, and using a geometric, first-order approach for describing gravity, in \cite{Olea:2005gb} it was shown that the Euler-Gauss-Bonnet term needed to regularize the four-dimensional gravitational action can be expressed, when written in well-adapted coordinates with radial direction orthogonal to the boundary, as a functional of the chosen boundary metric, extrinsic curvature and intrinsic curvature.
Then, in \cite{Olea:2005gb,Miskovic:2009bm}, it was shown that the same term provides the resummation of all the terms needed in the holographic renormalization scheme.

The four-dimensional $\mathcal{N}=2$ construction in \cite{Andrianopoli:2014aqa} has been 
reconsidered in \cite{Andrianopoli:2020zbl}, and an holographic framework for the four-dimensional $\mathcal{N}=2$ pure AdS$_4$ supergravity model, including all the contributions from the fermionic fields, was developed adopting a Fefferman-Graham parametrization, and working in the first-order formalism for the spin connection. The results of \cite{Andrianopoli:2020zbl} show that  the supersymmetric extension of the EGB term corresponding to the $\mathcal{N}=2$ boundary Lagrangian found in \cite{Andrianopoli:2014aqa},  when expressed in three-dimensional Fefferman-Graham intrinsic coordinates and expanded near the boundary along the direction orthogonal to it, precisely produces the counterterms necessary to  implement the holographic renormalization program at the full supersymmetric level.  It indeed  regularizes the effective action, exactly as the EGB term does for the bosonic case.

The analysis in \cite{Andrianopoli:2020zbl} extends to the $\mathcal{N}=2$ case previous results for the $\mathcal{N}=1$ case \cite{Amsel:2009rr}. Moreover,  the holographic analysis in \cite{Andrianopoli:2020zbl} includes all the components of the gravitini (the ones along the boundary, $\psi_\mu$, but also those orthogonal to it, $\psi_z$).
This was functional, in particular, as a step towards exploring a holographic description of the AVZ model at the quantum level.
Indeed, as shown in \cite{Andrianopoli:2018ymh},
the correspondence between AdS$_4$ $\mathcal{N}$-extended supergravity and the AVZ model requires to consider a torsionful version of the latter, based on the super-AdS$_3$ symmetry supergroup $\mathrm{OSp}(2|2)\times \mathrm{SO}(1,2)$. In this setup, the ansatz \eqref{avz} was  interpreted in terms of the condition projecting out the spin-$1/2$ component of the gravitino. Such condition, which is required to hold on-shell in the four-dimensional supergravity theory, reads
\begin{equation}\label{GammaPsizero}
    \Gamma^a V^{\hat\mu}_a\Psi_{\hat\mu}=0\,,
\end{equation} 
where, as before, $a=0,1,2,3$, $\hat\mu=0,1,2,3$ denote four-dimensional anholonomic and world indices, respectively, $\hat\Gamma^a $ here refer to gamma matrices in four dimensions, and $V^a$ is the bosonic vielbein in four-dimensions.
As discussed in \cite{Andrianopoli:2018ymh}, the ansatz \eqref{avz} follows by requiring that, in the dimensional reduction from four to three dimensions, the  relation \eqref{GammaPsizero} is realized non trivially, namely that it decomposes into
\begin{equation}\label{chiAVZ}
      \Gamma^i V^{\hat\mu}_i\Psi_{\hat\mu}= -\Gamma^3 V^{\hat\mu}_3\Psi_{\hat\mu}\propto \chi^{\text{(AVZ)}}    \,,
\end{equation}
where both terms in the decomposition have to be  non-vanishing (differently from what is general assumed in the holographic framework), being proportional to the dynamical spinor of the AVZ model.

The study in \cite{Andrianopoli:2020zbl} aimed to explore the relation between the classical local symmetries of the $\mathcal{N} = 2$ AdS$_4$ supergravity theory defined on the bulk and the quantum symmetries in the boundary field theory (namely, the \textit{asymptotic symmetries}, at radial infinity, of the gravitational background). The latter appear as residual symmetries left over after the gauge fixing of bulk local symmetries and whose parameters take value on the boundary. According to the observation in \eqref{chiAVZ}, to make contact with the AVZ model, in \cite{Andrianopoli:2020zbl} the holographic gauge fixing\footnote{The holographic gauge fixing consists in using the freedom on local parameters to fix the Lagrange
multipliers associated with the radial components of the fields when doing the holographic analysis.} was performed by refraining from imposing the standard condition $\Gamma^i V^{\hat\mu}_i\Psi_{\hat\mu}=0$  on the gravitino field at the boundary, in the reduction to three dimensions.

Ref. \cite{Andrianopoli:2020zbl} also includes a general discussion of the gauge fixing conditions on the bulk fields  yielding the asymptotic symmetries at the boundary. The corresponding currents of the boundary theory have been constructed and shown to satisfy the associated Ward identities, once the field equations of the bulk theory are imposed.
The asymptotic behaviour of the gravitini is determined by the \textit{supertorsion constraints}, associated with supersymmetry both in four- and three-dimensional spacetimes.
The near-boundary analysis of supergravity fields and local parameters was carried out in \cite{Andrianopoli:2020zbl} adopting the Fefferman-Graham parametrization for the metric and by performing the holographic gauge fixing of the bulk local symmetries, which provides the radial expansion of gauge fields and parameters. The residual transformations, leaving invariant the aforementioned gauge fixing, give rise to symmetry transformations on the boundary fields, associated with Noether currents conservation laws. Although in the first part of \cite{Andrianopoli:2020zbl} the multipliers in the holographic gauge fixing were fixed as generally as possible to pave the way for the possible holographic description of both the standard superconformal field theory (SCFT) and the AVZ model, the second part of the work was then centered on the standard SCFT case.

As the asymptotic symmetries of pure $\mathcal{N} = 2$ AdS$_4$ supergravity turn out to be given by the three-dimensional superconformal transformations, the latter are also asymptotic symmetries of an underlying boundary SCFT, according to the AdS/CFT correspondence, where the boundary fields act as sources for the corresponding operators in the dual SCFT. 
The superconformal  tranformations are associated with local parameters and  conserved currents (which are the quantum operators in the SCFT). The analysis of the quantum symmetries in a three-dimensional field theory holographically dual to $\mathcal{N} = 2$ AdS$_4$ supergravity was done in \cite{Andrianopoli:2020zbl} by explicitly deriving the expressions of the supercurrents and the corresponding conservation laws. The Ward identities expressing the above conservation laws on the supercurrents at the quantum level were then shown to hold off-shell in the boundary theory once the bulk field equations are imposed.

The boundary conditions on the supercurvatures are crucial to guarantee consistency of the holographic construction. In particular, the finiteness of the quantum generating
functional of the boundary theory was shown to require the vanishing on the boundary of the super-AdS curvatures, 
which was also proven, in \cite{Andrianopoli:2014aqa}, to be a necessary condition for a consistent definition of the bulk supergravity theory.

\vskip 5mm
Finally, let us comment on possible future applications of our formalism to the flat supergravity case, in a holographic context. 

As shortly reviewed in Section \ref{asympt}, the asymptotic symmetry of spacetime at null infinity is expected to be given by the infinite dimensional BMS$_4$ group.
On the other hand, in a holographic context a natural boundary dual to flat gravity has been recently identified in a seemingly unrelated theoretical framework, the one describing the so-called Carrollian fluids (see, e.g., \cite{Ciambelli:2018xat,Ciambelli:2018ojf,Campoleoni:2018ltl,Ciambelli:2018wre}). These are  fluids in 2 space dimensions whose dynamics  is left invariant   by the action of the infinite-dimensional ultra-relativistic conformal Carroll group.

Remarkably,  the BMS$_4$ and the (infinite-dimensional) conformal Carroll group were recently shown to be isomorphic \cite{Duval:2014uva}, thus strengthening the idea of the validity of a gauge/gravity holographic duality also for asymptotically flat spacetimes. 
Let us then briefly sketch in the following some group-theoretical motivations of this correspondence, following    \cite{Ciambelli:2018wre}, where  it was proven
that a holographic description of four-dimensional asymptotically locally flat spacetimes can be obtained  smoothly,  in a peculiar zero cosmological constant  limit of AdS holography. A (conformal) Carrollian geometry was shown to emerge in flat holography, since in the construction of \cite{Ciambelli:2018wre} the vanishing of the bulk cosmological constant   appears, from the boundary perspective,  as a  zero speed of light limit,   also known as Carrollian limit. 
More precisely, in \cite{Ciambelli:2018wre} the starting point is a four-dimensional bulk Einstein spacetime with negative cosmological constant $\Lambda = - 3 \kappa^2$, dual to a boundary relativistic fluid.
Then, the Ricci-flat limit is achieved in the limit $\kappa\rightarrow 0$ and, although no conformal boundary exists in this case, a \textit{two-dimensional spatial conformal structure emerges at null infinity}. As the Einstein bulk spacetime derivative expansion given in \cite{Ciambelli:2018wre} is performed along null tubes, it provides an appropriate setup to study both the nature of the two-dimensional spatial boundary and the
dynamics of the degrees of freedom it hosts as holographic duals to the bulk Ricci-flat
spacetime. Following \cite{Ciambelli:2018wre}, by parametrizing \textit{à la} Randers–Papapetrou the \textit{three-dimensional boundary spacetime metric}, that is
\begin{equation}\label{RandersPapapetrou}
\begin{split}
& \diff s^2 = - \kappa^2 \left( \mathbf{\Omega} \diff t - \mathbf{b}_{\mathfrak{a}} \diff x^{\mathfrak{a}} \right)^2 + \diff l^2  \,, \\
& \diff l^2 = \mathbf{a}_{\mathfrak{a}\mathfrak{b}} \diff x^{\mathfrak{a}} \diff x^{\mathfrak{b}} \,,
\end{split}
\end{equation}
where the tensors $\mathbf{\Omega}$, $\mathbf{b}_{\mathfrak{a}}$, $\mathbf{a}_{\mathfrak{a}\mathfrak{b}}$ are functions of all spacetime coordinates and $\diff l^2$ is the two-dimensional spatial metric,\footnote{Here we rename the two-dimensional indices $i,j,\dots$ of \cite{Ciambelli:2018wre} as $\mathfrak{a},\mathfrak{b},\ldots$, in order to avoid confusion with other indices adopted in the present work.} for vanishing $\kappa$ the time decouples in the boundary geometry. Here, let us mention that there actually exist two decoupling limits, associated with two inequivalent contractions of the Poincaré group: the Galilean (generally referred to as ``non-relativistic'') limit, which is formally obtained by sending the speed of light  to infinity, and the Carrollian (often called ``ultra-relativistic'') limit, which is formally reached sending the speed of light  to zero.\footnote{Let us warn the reader that sometimes the terminology ``non-relativistic'' is used in the literature to denote both Galilean and Carrollian structures,  none of  them being  relativistic, even if they emerge in totally different regimes.} Now, as in \eqref{RandersPapapetrou} $\kappa$ plays effectively the role of speed of light (while being proportional to the inverse of the AdS radius of the bulk asymptotic geometry), the flat limit $\kappa\rightarrow 0$ is precisely a Carrollian limit. This is the way in which a \textit{Carrollian boundary geometry} (i.e., covariant under \textit{Carrollian diffeomorphisms}), consisting of a spatial surface endowed with a positive-definite
metric $\diff l^2 = \mathbf{a}_{\mathfrak{a}\mathfrak{b}} \diff x^{\mathfrak{a}} \diff x^{\mathfrak{b}}$ and a Carrollian time $t \in \mathbb{R}$, emerges in this context. The Carrollian surface results to be the natural host for a \textit{conformal Carrollian fluid} (characterized by energy density and pressure related by a conformal equation of state, heat currents and traceless viscous stress tensor), which has to be considered as the holographic dual of a Ricci-flat spacetime,  its Carrollian fluid dynamics being dual to the gravitational bulk dynamics at the zero cosmological constant limit.
Conversely, as pointed out in \cite{Ciambelli:2018wre}, any Carrollian fluid evolving on a spatial surface with Carrollian geometry is associated with a Ricci-flat geometry. 
Flat holography is being the object of a large interest, and has been further investigated in various directions in the recent literature   \cite{Bagchi:2016bcd,Bagchi:2010zz,Bagchi:2012cy,Bagchi:2019xfx}, mostly at the bosonic level.

A supersymmetric extension of the duality encoded in flat holography was poorly investigated so far.  
More specifically,  referring to  the formalism reviewed in the previous sections of the present paper,  its  possible application to flat holography is still an open problem. 
In Section \ref{flat} we reviewed how the supersymmetry invariance of flat supergravity with boundary can be recovered by inclusion of 1-form auxiliary fields, $A^{ab}$ and $\chi$, contributing to the boundary Lagrangian. This is independent from the location of the boundary itself, 
and the same auxiliary fields  are then  expected to play a crucial role in the application of our formalism to a flat-holography gauge/gravity correspondence. 
Furthermore, since the  boundary symmetry of $D=4$ flat supergravity was found in \cite{Concha:2018ywv} to be the super-Maxwell algebra, as  reviewed in Section \ref{flat}, we expect a relation to emerge, for the case of an asymptotic boundary,  between the super-Maxwell algebra and the asymptotic boundary symmetry of asymptotically flat (super)space, that is  super-BMS$_4$ algebra (or, equivalently, the super-Carroll algebra, see, e.g., \cite{Bergshoeff:2015wma,Ravera:2019ize,Ali:2019jjp}).

Verification of our hypotesis would require, as a first step, to perform an intrinsic description of the boundary Lagrangian for the case of a null boundary geometry, and the corresponding decomposition of the tensorial structures with respect to those covariant with respect to the symmetries on the chosen boundary.
This is left to future investigation.

\section*{Acknowledgements}

L.R. would like to thank the Department of Applied Science and Technology of the Polytechnic of Turin for financial support.


\begin{thebibliography}{199}

\bibitem{York:1972sj}
J.~W.~York, Jr.,
``Role of conformal three geometry in the dynamics of gravitation,''
Phys. Rev. Lett. \textbf{28} (1972), 1082-1085
doi:10.1103/PhysRevLett.28.1082

\bibitem{Gibbons:1976ue}
G.~W.~Gibbons and S.~W.~Hawking,
``Action Integrals and Partition Functions in Quantum Gravity,''
Phys. Rev. D \textbf{15} (1977), 2752-2756
doi:10.1103/PhysRevD.15.2752

\bibitem{Aros:1999id}
R.~Aros, M.~Contreras, R.~Olea, R.~Troncoso and J.~Zanelli,
``Conserved charges for gravity with locally AdS asymptotics,''
Phys. Rev. Lett. \textbf{84} (2000), 1647-1650
doi:10.1103/PhysRevLett.84.1647
[arXiv:gr-qc/9909015 [gr-qc]].

\bibitem{Aros:1999kt}
R.~Aros, M.~Contreras, R.~Olea, R.~Troncoso and J.~Zanelli,
``Conserved charges for even dimensional asymptotically AdS gravity theories,''
Phys. Rev. D \textbf{62} (2000), 044002
doi:10.1103/PhysRevD.62.044002
[arXiv:hep-th/9912045 [hep-th]].

\bibitem{Mora:2004kb}
P.~Mora, R.~Olea, R.~Troncoso and J.~Zanelli,
``Finite action principle for Chern-Simons AdS gravity,''
JHEP \textbf{06} (2004), 036
doi:10.1088/1126-6708/2004/06/036
[arXiv:hep-th/0405267 [hep-th]].

\bibitem{Olea:2005gb}
R.~Olea,
``Mass, angular momentum and thermodynamics in four-dimensional Kerr-AdS black holes,''
JHEP \textbf{06} (2005), 023
doi:10.1088/1126-6708/2005/06/023
[arXiv:hep-th/0504233 [hep-th]].

\bibitem{Jatkar:2014npa}
D.~P.~Jatkar, G.~Kofinas, O.~Miskovic and R.~Olea,
``Conformal Mass in AdS gravity,''
Phys. Rev. D \textbf{89} (2014) no.12, 124010
doi:10.1103/PhysRevD.89.124010
[arXiv:1404.1411 [hep-th]].

\bibitem{Jatkar:2015ffa}
D.~P.~Jatkar, G.~Kofinas, O.~Miskovic and R.~Olea, \\
``Conformal mass in Einstein-Gauss-Bonnet AdS gravity,''
Phys. Rev. D \textbf{91} (2015) no.10, 105030
doi:10.1103/PhysRevD.91.105030
[arXiv:1501.06861 [hep-th]].

\bibitem{Belyaev:2005rt}
D.~V.~Belyaev,
``Boundary conditions in supergravity on a manifold with boundary,''
JHEP {\bf 0601} (2006) 047
[hep-th/0509172].
  
\bibitem{Belyaev:2007bg}
D.~V.~Belyaev and P.~van Nieuwenhuizen,
``Tensor calculus for supergravity on a manifold with boundary,''
JHEP {\bf 0802} (2008) 047
[arXiv:0711.2272 [hep-th]].

\bibitem{Belyaev:2008ex}
D.~V.~Belyaev and P.~van Nieuwenhuizen,
``Simple d=4 supergravity with a boundary,''
JHEP \textbf{09} (2008), 069
doi:10.1088/1126-6708/2008/09/069
[arXiv:0806.4723 [hep-th]].

\bibitem{Grumiller:2009dx}
D.~Grumiller and P.~van Nieuwenhuizen,
``Holographic counterterms from local supersymmetry without boundary conditions,''
Phys.\ Lett.\ B {\bf 682} (2010) 462
[arXiv:0908.3486 [hep-th]].

\bibitem{Belyaev:2010as}
D.~V.~Belyaev and T.~G.~Pugh,
``The Supermultiplet of boundary conditions in supergravity,''
JHEP {\bf 1010} (2010) 031
[arXiv:1008.1574 [hep-th]].

\bibitem{Maldacena:1997re}
J.~M.~Maldacena,
``The Large N limit of superconformal field theories and supergravity,''
Adv. Theor. Math. Phys. \textbf{2} (1998), 231-252
doi:10.1023/A:1026654312961
[arXiv:hep-th/9711200 [hep-th]].

\bibitem{Gubser:1998bc}
S.~S.~Gubser, I.~R.~Klebanov and A.~M.~Polyakov,
``Gauge theory correlators from noncritical string theory,''
Phys. Lett. B \textbf{428} (1998), 105-114
doi:10.1016/S0370-2693(98)00377-3
[arXiv:hep-th/9802109 [hep-th]].

\bibitem{Witten:1998qj}
E.~Witten,
``Anti-de Sitter space and holography,''
Adv. Theor. Math. Phys. \textbf{2} (1998), 253-291
doi:10.4310/ATMP.1998.v2.n2.a2
[arXiv:hep-th/9802150 [hep-th]].

\bibitem{Aharony:1999ti}
O.~Aharony, S.~S.~Gubser, J.~M.~Maldacena, H.~Ooguri and Y.~Oz,
``Large N field theories, string theory and gravity,''
Phys. Rept. \textbf{323} (2000), 183-386
doi:10.1016/S0370-1573(99)00083-6
[arXiv:hep-th/9905111 [hep-th]].

\bibitem{Balasubramanian:1999re}
V.~Balasubramanian and P.~Kraus,
``A Stress tensor for Anti-de Sitter gravity,''
Commun. Math. Phys. \textbf{208} (1999), 413-428
doi:10.1007/s002200050764
[arXiv:hep-th/9902121 [hep-th]].

\bibitem{deBoer:1999tgo}
J.~de Boer, E.~P.~Verlinde and H.~L.~Verlinde,
``On the holographic renormalization group,''
JHEP \textbf{08} (2000), 003
doi:10.1088/1126-6708/2000/08/003
[arXiv:hep-th/9912012 [hep-th]].

\bibitem{Verlinde:1999xm}
E.~P.~Verlinde and H.~L.~Verlinde,
``RG flow, gravity and the cosmological constant,''
JHEP \textbf{05} (2000), 034
doi:10.1088/1126-6708/2000/05/034
[arXiv:hep-th/9912018 [hep-th]].

\bibitem{deBoer:2000cz}
J.~de Boer, \\
``The Holographic renormalization group,''
Fortsch. Phys. \textbf{49} (2001), 339-358
doi:10.1002/1521-3978(200105)49:4/6\ensuremath{<}339::AID-PROP339\ensuremath{>}3.0.CO;2-A
[arXiv:hep-th/0101026 [hep-th]].

\bibitem{deHaro:2000vlm}
S.~de Haro, S.~N.~Solodukhin and K.~Skenderis,
``Holographic reconstruction of space-time and renormalization in the AdS / CFT correspondence,''
Commun. Math. Phys. \textbf{217} (2001), 595-622
doi:10.1007/s002200100381
[arXiv:hep-th/0002230 [hep-th]].

\bibitem{Skenderis:2002wp}
K.~Skenderis,
``Lecture notes on holographic renormalization,''
Class. Quant. Grav. \textbf{19} (2002), 5849-5876
doi:10.1088/0264-9381/19/22/306
[arXiv:hep-th/0209067 [hep-th]].

\bibitem{Miskovic:2009bm}
O.~Miskovic and R.~Olea, \\
``Topological regularization and self-duality in four-dimensional anti-de Sitter gravity,''
Phys. Rev. D \textbf{79} (2009), 124020
doi:10.1103/PhysRevD.79.124020
[arXiv:0902.2082 [hep-th]].

\bibitem{Anastasiou:2020zwc}
G.~Anastasiou, O.~Miskovic, R.~Olea and I.~Papadimitriou,
``Counterterms, Kounterterms, and the variational problem in AdS gravity,''
JHEP \textbf{08} (2020), 061
doi:10.1007/JHEP08(2020)061
[arXiv:2003.06425 [hep-th]].

\bibitem{MacDowell:1977jt}
S.~W.~MacDowell and F.~Mansouri,
``Unified Geometric Theory of Gravity and Supergravity,''
Phys. Rev. Lett. \textbf{38} (1977), 739
[erratum: Phys. Rev. Lett. \textbf{38} (1977), 1376]
doi:10.1103/PhysRevLett.38.739

\bibitem{Amsel:2009rr}
A.~J.~Amsel and G.~Compere,
``Supergravity at the boundary of AdS supergravity,''
Phys. Rev. D \textbf{79} (2009), 085006
[arXiv:0901.3609 [hep-th]].

\bibitem{vanNieuwenhuizen:2005kg}
P.~van Nieuwenhuizen and D.~V.~Vassilevich,
``Consistent boundary conditions for supergravity,''
Class. Quant. Grav. \textbf{22} (2005), 5029-5051
doi:10.1088/0264-9381/22/23/008
[arXiv:hep-th/0507172 [hep-th]].

\bibitem{Esposito:1996hu}
G.~Esposito, A.~Y.~Kamenshchik and K.~Kirsten,
``One loop effective action for Euclidean Maxwell theory on manifolds with boundary,''
Phys. Rev. D \textbf{54} (1996), 7328-7337
doi:10.1103/PhysRevD.54.7328
[arXiv:hep-th/9606132 [hep-th]].

\bibitem{Avramidi:1997hy}
I.~G.~Avramidi and G.~Esposito,
``Gauge theories on manifolds with boundary,''
Commun. Math. Phys. \textbf{200} (1999), 495-543
doi:10.1007/s002200050539
[arXiv:hep-th/9710048 [hep-th]].

\bibitem{Moss:2003bk}
I.~G.~Moss,
``Boundary terms for eleven-dimensional supergravity and M theory,''
Phys. Lett. B \textbf{577} (2003), 71-75
doi:10.1016/j.physletb.2003.10.027
[arXiv:hep-th/0308159 [hep-th]].

\bibitem{Moss:2004ck}
I.~G.~Moss,
``Boundary terms for supergravity and heterotic M theory,''
Nucl. Phys. B \textbf{729} (2005), 179-202
doi:10.1016/j.nuclphysb.2005.09.023
[arXiv:hep-th/0403106 [hep-th]].

\bibitem{Howe:2011tm}
P.~S.~Howe, T.~G.~Pugh, K.~S.~Stelle and C.~Strickland-Constable,
``Ectoplasm with an Edge,''
JHEP \textbf{08} (2011), 081
doi:10.1007/JHEP08(2011)081
[arXiv:1104.4387 [hep-th]].

\bibitem{Andrianopoli:2014aqa}
L.~Andrianopoli and R.~D'Auria,
``N=1 and N=2 pure supergravities on a manifold with boundary,''
JHEP \textbf{08} (2014), 012
doi:10.1007/JHEP08(2014)012
[arXiv:1405.2010 [hep-th]].

\bibitem{Castellani:1991et}
L.~Castellani, R.~D'Auria and P.~Fré,
``Supergravity and superstrings: A Geometric perspective. Vol. 1 and 2,'' Published in Singapore: World Scientific.

\bibitem{DAuria:2020guc}
R.~D'Auria,
``Geometric supergravitty,''
Review article from the book \textit{``Tullio Regge: An Eclectic Genius - From Quantum Gravity to Computer Play''}, World Scientific Publishing Co. Pte. Ltd. (2019),
[arXiv:2005.13593 [hep-th]].

\bibitem{Ipinza:2016con}
P.~K.~Concha, M.~C.~Ipinza, L.~Ravera and E.~K.~Rodr\'\i{}guez,
``On the Supersymmetric Extension of Gauss-Bonnet like Gravity,''
JHEP \textbf{09} (2016), 007
doi:10.1007/JHEP09(2016)007
[arXiv:1607.00373 [hep-th]].

\bibitem{Banaudi:2018zmh}
A.~Banaudi and L.~Ravera,
``Generalized AdS-Lorentz deformed supergravity on a manifold with boundary,''
Eur. Phys. J. Plus \textbf{133} (2018) no.12, 514
doi:10.1140/epjp/i2018-12335-0
[arXiv:1803.08738 [hep-th]].

\bibitem{deAzcarraga:2010sw}
J.~A.~de Azcarraga, K.~Kamimura and J.~Lukierski, \\
``Generalized cosmological term from Maxwell symmetries,''
Phys. Rev. D \textbf{83} (2011), 124036
doi:10.1103/PhysRevD.83.124036
[arXiv:1012.4402 [hep-th]].

\bibitem{Soroka:2011tc}
D.~V.~Soroka and V.~A.~Soroka,
``Gauge semi-simple extension of the Poincar\'e group,''
Phys. Lett. B \textbf{707} (2012), 160-162
doi:10.1016/j.physletb.2011.07.003
[arXiv:1101.1591 [hep-th]].

\bibitem{Durka:2011gm}
R.~Durka, J.~Kowalski-Glikman and M.~Szczachor,
``AdS-Maxwell superalgebra and supergravity,''
Mod. Phys. Lett. A \textbf{27} (2012), 1250023
doi:10.1142/S021773231250023X
[arXiv:1107.5731 [hep-th]].

\bibitem{Cebecioglu:2014rca}
O.~Cebecio\u{g}lu and S.~Kibaro\u{g}lu,
``Gauge theory of the Maxwell-Weyl group,''
Phys. Rev. D \textbf{90} (2014) no.8, 084053
doi:10.1103/PhysRevD.90.084053
[arXiv:1404.3969 [hep-th]].

\bibitem{Concha:2015tla}
P.~K.~Concha, E.~K.~Rodr\'\i{}guez and P.~Salgado,
``Generalized supersymmetric cosmological term in $N=1$ Supergravity,''
JHEP \textbf{08} (2015), 009
doi:10.1007/JHEP08(2015)009
[arXiv:1504.01898 [hep-th]].

\bibitem{Penafiel:2018vpe}
D.~M.~Pe\~nafiel and L.~Ravera,
``Generalized cosmological term in $D=4$ supergravity from a new AdS\textendash{}Lorentz superalgebra,''
Eur. Phys. J. C \textbf{78} (2018) no.11, 945
doi:10.1140/epjc/s10052-018-6421-9
[arXiv:1807.07673 [hep-th]].

\bibitem{Kibaroglu:2018oue}
S.~Kibaro\u{g}lu and O.~Cebecio\u{g}lu,
``$D=4$ supergravity from the Maxwell-Weyl superalgebra,''
Eur. Phys. J. C \textbf{79} (2019) no.11, 898
doi:10.1140/epjc/s10052-019-7421-0
[arXiv:1812.09861 [hep-th]].

\bibitem{Eder:2021nyb}
K.~Eder and H.~Sahlmann,
``Holst-MacDowell-Mansouri action for (extended) supergravity with boundaries and super Chern-Simons theory,''
[arXiv:2104.02011 [gr-qc]].

\bibitem{Hojman:1980kv}
R.~Hojman, C.~Mukku and W.~A.~Sayed,
``Parity violation in metric-torsion theories of gravitation,''
Phys. Rev. D \textbf{22} (1980), 1915-1921
doi:10.1103/PhysRevD.22.1915

\bibitem{Iosifidis:2020dck}
D.~Iosifidis and L.~Ravera, \\
``Parity Violating Metric-Affine Gravity Theories,''
Class. Quant. Grav. \textbf{38} (2021) no.11, 115003
doi:10.1088/1361-6382/abde1a
[arXiv:2009.03328 [gr-qc]].

\bibitem{Holst:1995pc}
S.~Holst,
``Barbero's Hamiltonian derived from a generalized Hilbert-Palatini action,''
Phys. Rev. D \textbf{53} (1996), 5966-5969
doi:10.1103/PhysRevD.53.5966
[arXiv:gr-qc/9511026 [gr-qc]].

\bibitem{vanNieuwenhuizen:2006pz}
P.~van Nieuwenhuizen, A.~Rebhan, D.~V.~Vassilevich and R.~Wimmer,
``Boundary terms in supergravity and supersymmetry,''
Int. J. Mod. Phys. D \textbf{15} (2006), 1643-1658
doi:10.1142/S0218271806009017
[arXiv:hep-th/0606075 [hep-th]].

\bibitem{Concha:2018ywv}
P.~Concha, L.~Ravera and E.~Rodr\'\i{}guez,
``On the supersymmetry invariance of flat supergravity with boundary,''
JHEP \textbf{01} (2019), 192
doi:10.1007/JHEP01(2019)192
[arXiv:1809.07871 [hep-th]].

\bibitem{Bonanos:2009wy}
S.~Bonanos, J.~Gomis, K.~Kamimura and J.~Lukierski,
``Maxwell Superalgebra and Superparticle in Constant Gauge Backgrounds,''
Phys. Rev. Lett. \textbf{104} (2010), 090401
doi:10.1103/PhysRevLett.104.090401
[arXiv:0911.5072 [hep-th]].

\bibitem{Concha:2014tca}
P.~K.~Concha and E.~K.~Rodr\'\i{}guez,
``N = 1 Supergravity and Maxwell superalgebras,''
JHEP \textbf{09} (2014), 090
doi:10.1007/JHEP09(2014)090
[arXiv:1407.4635 [hep-th]].

\bibitem{Soroka:2006aj}
D.~V.~Soroka and V.~A.~Soroka,
``Semi-simple extension of the (super)Poincaré algebra,''
Adv. High Energy Phys. \textbf{2009} (2009), 234147
doi:10.1155/2009/234147
[arXiv:hep-th/0605251 [hep-th]].

\bibitem{Gomis:2009dm}
J.~Gomis, K.~Kamimura and J.~Lukierski,
``Deformations of Maxwell algebra and their Dynamical Realizations,''
JHEP \textbf{08} (2009), 039
doi:10.1088/1126-6708/2009/08/039
[arXiv:0906.4464 [hep-th]].

\bibitem{Diaz:2012zza}
J.~Diaz, O.~Fierro, F.~Izaurieta, N.~Merino, E.~Rodriguez, P.~Salgado and O.~Valdivia,
``A generalized action for (2 + 1)-dimensional Chern-Simons gravity,''
J. Phys. A \textbf{45} (2012), 255207
doi:10.1088/1751-8113/45/25/255207
[arXiv:1311.2215 [gr-qc]].

\bibitem{Salgado:2013eut}
P.~Salgado and S.~Salgado,
``$\mathfrak{so}(D-1,1)\otimes \mathfrak{so}(D-1,2)$ algebras and gravity,''
Phys. Lett. B \textbf{728} (2014), 5-10
doi:10.1016/j.physletb.2013.11.009

\bibitem{Schrader:1972zd}
R.~Schrader,
``The maxwell group and the quantum theory of particles in classical homogeneous electromagnetic fields,''
Fortsch. Phys. \textbf{20} (1972), 701-734
doi:10.1002/prop.19720201202

\bibitem{DAuria:1982uck}
R.~D'Auria and P.~Fré,
``Geometric Supergravity in d = 11 and Its Hidden Supergroup,''
Nucl. Phys. B \textbf{201} (1982), 101-140
[erratum: Nucl. Phys. B \textbf{206} (1982), 496]
doi:10.1016/0550-3213(82)90281-4

\bibitem{Castellani:1982kd}
L.~Castellani, P.~Fre, F.~Giani, K.~Pilch and P.~van Nieuwenhuizen,
``Gauging of $d=11$ Supergravity?,''
Annals Phys. \textbf{146} (1983), 35
doi:10.1016/0003-4916(83)90052-0

\bibitem{Green:1989nn}
M.~B.~Green,
``Supertranslations, Superstrings and {Chern-Simons} Forms,''
Phys. Lett. B \textbf{223} (1989), 157-164
doi:10.1016/0370-2693(89)90233-5

\bibitem{Andrianopoli:2016osu}
L.~Andrianopoli, R.~D'Auria and L.~Ravera,
``Hidden Gauge Structure of Supersymmetric Free Differential Algebras,''
JHEP \textbf{08} (2016), 095
doi:10.1007/JHEP08(2016)095
[arXiv:1606.07328 [hep-th]].

\bibitem{Andrianopoli:2017itj}
L.~Andrianopoli, R.~D'Auria and L.~Ravera,
``More on the Hidden Symmetries of 11D Supergravity,''
Phys. Lett. B \textbf{772} (2017), 578-585
doi:10.1016/j.physletb.2017.07.016
[arXiv:1705.06251 [hep-th]].

\bibitem{Penafiel:2017wfr}
D.~M.~Pe\~nafiel and L.~Ravera,
``On the Hidden Maxwell Superalgebra underlying D=4 Supergravity,''
Fortsch. Phys. \textbf{65} (2017) no.9, 1700005
doi:10.1002/prop.201700005
[arXiv:1701.04234 [hep-th]].

\bibitem{Ravera:2018vra}
L.~Ravera,
``Hidden role of Maxwell superalgebras in the free differential algebras of D = 4 and D = 11 supergravity,''
Eur. Phys. J. C \textbf{78} (2018) no.3, 211
doi:10.1140/epjc/s10052-018-5673-8
[arXiv:1801.08860 [hep-th]].

\bibitem{Neeman:1978njh}
Y.~Ne'eman and T.~Regge,
``Gauge Theory of Gravity and Supergravity on a Group Manifold,''
Riv. Nuovo Cim. \textbf{1N5} (1978), 1
doi:10.1007/BF02724472

\bibitem{Neeman:1978zvv}
Y.~Ne'eman and T.~Regge,
``Gravity and Supergravity as Gauge Theories on a Group Manifold,''
Phys. Lett. B \textbf{74} (1978), 54-56
doi:10.1016/0370-2693(78)90058-8

\bibitem{DAuria:2021dth}
R.~D'Auria and L.~Ravera,
``(Super)conformal gravity with totally antisymmetric torsion,''
[arXiv:2101.10978 [hep-th]].

\bibitem{Townsend:1977qa}
P.~K.~Townsend,
``Cosmological Constant in Supergravity,''
Phys. Rev. D \textbf{15} (1977), 2802-2804
doi:10.1103/PhysRevD.15.2802

\bibitem{Fayet:1974jb}
P.~Fayet and J.~Iliopoulos,
``Spontaneously Broken Supergauge Symmetries and Goldstone Spinors,''
Phys. Lett. B \textbf{51} (1974), 461-464
doi:10.1016/0370-2693(74)90310-4

\bibitem{Castellani:2014goa}
L.~Castellani, R.~Catenacci and P.~A.~Grassi,
``Supergravity Actions with Integral Forms,''
Nucl. Phys. B \textbf{889} (2014), 419-442
doi:10.1016/j.nuclphysb.2014.10.023
[arXiv:1409.0192 [hep-th]].

\bibitem{Castellani:2015paa}
L.~Castellani, R.~Catenacci and P.~A.~Grassi,
``The Geometry of Supermanifolds and New Supersymmetric Actions,''
Nucl. Phys. B \textbf{899} (2015), 112-148
doi:10.1016/j.nuclphysb.2015.07.028
[arXiv:1503.07886 [hep-th]].

\bibitem{Andrianopoli:1996cm}
L.~Andrianopoli, M.~Bertolini, A.~Ceresole, R.~D'Auria, S.~Ferrara, P.~Fré and T.~Magri,
``N=2 supergravity and N=2 superYang-Mills theory on general scalar manifolds: Symplectic covariance, gaugings and the momentum map,''
J. Geom. Phys. \textbf{23} (1997), 111-189
doi:10.1016/S0393-0440(97)00002-8
[arXiv:hep-th/9605032 [hep-th]].

\bibitem{Andrianopoli:2020zbl}
L.~Andrianopoli, B.~L.~Cerchiai, R.~Matrecano, O.~Miskovic, R.~Noris, R.~Olea, L.~Ravera and M.~Trigiante,
``$\mathcal{N}$ = 2 AdS$_{4}$ supergravity, holography and Ward identities,''
JHEP \textbf{02} (2021), 141
doi:10.1007/JHEP02(2021)141
[arXiv:2010.02119 [hep-th]].

\bibitem{Andrianopoli:2019sip}
L.~Andrianopoli, B.~L.~Cerchiai, R.~D'Auria, A.~Gallerati, R.~Noris, M.~Trigiante and J.~Zanelli,
``$\mathcal{N}$-extended $D = 4$ supergravity, unconventional SUSY and graphene,''
JHEP \textbf{01} (2020), 084
doi:10.1007/JHEP01(2020)084
[arXiv:1910.03508 [hep-th]].

\bibitem{Bondi:1962px}
H.~Bondi, M.~G.~J.~van der Burg and A.~W.~K.~Metzner,
``Gravitational waves in general relativity. 7. Waves from axisymmetric isolated systems,''
Proc. Roy. Soc. Lond. A \textbf{269} (1962), 21-52
doi:10.1098/rspa.1962.0161

\bibitem{Sachs:1962}
R.~Sachs,
``Gravitational waves in general relativity VIII. Waves in asymptotically flat space-time,''
Proc. R. Soc. Lond. A \textbf{270}, 103–126
doi:10.1098/rspa.1962.0206

\bibitem{Sachs:1962zza}
R.~Sachs,
``Asymptotic symmetries in gravitational theory,''
Phys. Rev. \textbf{128} (1962), 2851-2864
doi:10.1103/PhysRev.128.2851

\bibitem{Barnich:2009se}
G.~Barnich and C.~Troessaert,
``Symmetries of asymptotically flat 4 dimensional spacetimes at null infinity revisited,''
Phys. Rev. Lett. \textbf{105} (2010), 111103
doi:10.1103/PhysRevLett.105.111103
[arXiv:0909.2617 [gr-qc]].

\bibitem{Barnich:2010eb}
G.~Barnich and C.~Troessaert,
``Aspects of the BMS/CFT correspondence,''
JHEP \textbf{05} (2010), 062
doi:10.1007/JHEP05(2010)062
[arXiv:1001.1541 [hep-th]].

\bibitem{Barnich:2011mi}
G.~Barnich and C.~Troessaert,
``BMS charge algebra,''
JHEP \textbf{12} (2011), 105
doi:10.1007/JHEP12(2011)105
[arXiv:1106.0213 [hep-th]].

\bibitem{Barnich:2013axa}
G.~Barnich and C.~Troessaert,
``Comments on holographic current algebras and asymptotically flat four dimensional spacetimes at null infinity,''
JHEP \textbf{11} (2013), 003
doi:10.1007/JHEP11(2013)003
[arXiv:1309.0794 [hep-th]].

\bibitem{Strominger:2013jfa}
A.~Strominger,
``On BMS Invariance of Gravitational Scattering,''
JHEP \textbf{07} (2014), 152
doi:10.1007/JHEP07(2014)152
[arXiv:1312.2229 [hep-th]].

\bibitem{Duval:2014uva}
C.~Duval, G.~W.~Gibbons and P.~A.~Horvathy,
``Conformal Carroll groups and BMS symmetry,''
Class. Quant. Grav. \textbf{31} (2014), 092001
doi:10.1088/0264-9381/31/9/092001
[arXiv:1402.5894 [gr-qc]].

\bibitem{Barnich:2016lyg}
G.~Barnich and C.~Troessaert,
``Finite BMS transformations,''
JHEP \textbf{03} (2016), 167
doi:10.1007/JHEP03(2016)167
[arXiv:1601.04090 [gr-qc]].

\bibitem{Bagchi:2016bcd}
A.~Bagchi, R.~Basu, A.~Kakkar and A.~Mehra,
``Flat Holography: Aspects of the dual field theory,''
JHEP \textbf{12} (2016), 147
doi:10.1007/JHEP12(2016)147
[arXiv:1609.06203 [hep-th]].

\bibitem{Bagchi:2019clu}
A.~Bagchi, R.~Basu, A.~Mehra and P.~Nandi,
``Field Theories on Null Manifolds,''
JHEP \textbf{02} (2020), 141
doi:10.1007/JHEP02(2020)141
[arXiv:1912.09388 [hep-th]].

\bibitem{Aneesh:2021uzk}
P.~B.~Aneesh, G.~Comp\`ere, L.~P.~de Gioia, I.~Mol and B.~Swidler,
``Celestial Holography: Lectures on Asymptotic Symmetries,''
[arXiv:2109.00997 [hep-th]].

\bibitem{Gupta:2021cwo}
N.~Gupta, P.~Paul and N.~V.~Suryanarayana,
``An $\widehat{sl_2}$ Symmetry of ${\mathbb R}^{1,3}$ Gravity,''
[arXiv:2109.06857 [hep-th]].

\bibitem{Brown:1986nw}
J.~D.~Brown and M.~Henneaux,
``Central Charges in the Canonical Realization of Asymptotic Symmetries: An Example from Three-Dimensional Gravity,''
Commun. Math. Phys. \textbf{104} (1986), 207-226
doi:10.1007/BF01211590

\bibitem{Frauendiener:2006}
J.~Frauendiener,
``Asymptotic Structure and Conformal Infinity,''
Encyclopedia of Mathematical Physics,
Academic Press (2006), 221-226,
ISBN 9780125126663,
doi:10.1016/B0-12-512666-2/00011-0

\bibitem{Flanagan:2015pxa}
\'E.~\'E.~Flanagan and D.~A.~Nichols,
``Conserved charges of the extended Bondi-Metzner-Sachs algebra,''
Phys. Rev. D \textbf{95} (2017) no.4, 044002
doi:10.1103/PhysRevD.95.044002
[arXiv:1510.03386 [hep-th]].

\bibitem{Strominger:2017zoo}
A.~Strominger,
``Lectures on the Infrared Structure of Gravity and Gauge Theory,''
[arXiv:1703.05448 [hep-th]].

\bibitem{Pasterski:2016qvg}
S.~Pasterski, S.~H.~Shao and A.~Strominger,
``Flat Space Amplitudes and Conformal Symmetry of the Celestial Sphere,''
Phys. Rev. D \textbf{96} (2017) no.6, 065026
doi:10.1103/PhysRevD.96.065026
[arXiv:1701.00049 [hep-th]].

\bibitem{Arkani-Hamed:2020gyp}
N.~Arkani-Hamed, M.~Pate, A.~M.~Raclariu and A.~Strominger,
``Celestial amplitudes from UV to IR,''
JHEP \textbf{08} (2021), 062
doi:10.1007/JHEP08(2021)062
[arXiv:2012.04208 [hep-th]].

\bibitem{LevyLeblond:1965}
J.~Levy-Leblond, ``Une nouvelle limite non-relativiste du group de Poincaré'' (in French),
Ann. Inst. H. Poincaré \textbf{3} (1965) 1.

\bibitem{Awada:1985by}
M.~A.~Awada, G.~W.~Gibbons and W.~T.~Shaw,
``CONFORMAL SUPERGRAVITY, TWISTORS AND THE SUPER BMS GROUP,''
Annals Phys. \textbf{171} (1986), 52
doi:10.1016/S0003-4916(86)80023-9

\bibitem{Henneaux:2020ekh}
M.~Henneaux, J.~Matulich and T.~Neogi, \\
``Asymptotic realization of the super-BMS algebra at spatial infinity,''
Phys. Rev. D \textbf{101} (2020) no.12, 126016
doi:10.1103/PhysRevD.101.126016
[arXiv:2004.07299 [hep-th]].

\bibitem{Avery:2015iix}
S.~G.~Avery and B.~U.~W.~Schwab,
``Residual Local Supersymmetry and the Soft Gravitino,''
Phys. Rev. Lett. \textbf{116} (2016) no.17, 171601
doi:10.1103/PhysRevLett.116.171601
[arXiv:1512.02657 [hep-th]].

\bibitem{Fotopoulos:2020bqj}
A.~Fotopoulos, S.~Stieberger, T.~R.~Taylor and B.~Zhu,
``Extended Super BMS Algebra of Celestial CFT,''
JHEP \textbf{09} (2020), 198
doi:10.1007/JHEP09(2020)198
[arXiv:2007.03785 [hep-th]].

\bibitem{Narayanan:2020amh}
S.~A.~Narayanan, \\
``Massive Celestial Fermions,''
JHEP \textbf{12} (2020), 074
doi:10.1007/JHEP12(2020)074
[arXiv:2009.03883 [hep-th]].

\bibitem{Fuentealba:2021xhn}
O.~Fuentealba, M.~Henneaux, S.~Majumdar, J.~Matulich and T.~Neogi,
``Local supersymmetry and the square roots of Bondi-Metzner-Sachs supertranslations,''
[arXiv:2108.07825 [hep-th]].

\bibitem{Avery:2015gxa}
S.~G.~Avery and B.~U.~W.~Schwab,
``Burg-Metzner-Sachs symmetry, string theory, and soft theorems,''
Phys. Rev. D \textbf{93} (2016), 026003
doi:10.1103/PhysRevD.93.026003
[arXiv:1506.05789 [hep-th]].

\bibitem{Bacry:1970ye}
H.~Bacry, P.~Combe and J.~L.~Richard,
``Group-theoretical analysis of elementary particles in an external electromagnetic field. 1. the relativistic particle in a constant and uniform field,''
Nuovo Cim. A \textbf{67} (1970), 267-299
doi:10.1007/BF02725178

\bibitem{Concha:2015woa}
P.~K.~Concha, O.~Fierro, E.~K.~Rodr\'\i{}guez and P.~Salgado,
``Chern\textendash{}Simons supergravity in D=3 and Maxwell superalgebra,''
Phys. Lett. B \textbf{750} (2015), 117-121
doi:10.1016/j.physletb.2015.09.005
[arXiv:1507.02335 [hep-th]].

\bibitem{Concha:2018jxx}
P.~Concha, D.~M.~Pe\~nafiel and E.~Rodr\'\i{}guez, \\
``On the Maxwell supergravity and flat limit in 2 + 1 dimensions,''
Phys. Lett. B \textbf{785} (2018), 247-253
doi:10.1016/j.physletb.2018.08.050
[arXiv:1807.00194 [hep-th]].

\bibitem{deAzcarraga:2012zv}
J.~A.~de Azcarraga, J.~M.~Izquierdo, J.~Lukierski and M.~Woronowicz,
``Generalizations of Maxwell (super)algebras by the expansion method,''
Nucl. Phys. B \textbf{869} (2013), 303-314
doi:10.1016/j.nuclphysb.2012.12.008
[arXiv:1210.1117 [hep-th]].

\bibitem{deAzcarraga:2014jpa}
J.~A.~de Azcarraga and J.~M.~Izquierdo, \\
``Minimal D = 4 supergravity from the superMaxwell algebra,''
Nucl. Phys. B \textbf{885} (2014), 34-45
doi:10.1016/j.nuclphysb.2014.05.007
[arXiv:1403.4128 [hep-th]].

\bibitem{Concha:2014xfa}
P.~K.~Concha and E.~K.~Rodr\'\i{}guez,
``Maxwell Superalgebras and Abelian Semigroup Expansion,''
Nucl. Phys. B \textbf{886} (2014), 1128-1152
doi:10.1016/j.nuclphysb.2014.07.022
[arXiv:1405.1334 [hep-th]].

\bibitem{Izaurieta:2006zz}
F.~Izaurieta, E.~Rodriguez and P.~Salgado,
``Expanding Lie (super)algebras through Abelian semigroups,''
J. Math. Phys. \textbf{47} (2006), 123512
doi:10.1063/1.2390659
[arXiv:hep-th/0606215 [hep-th]].

\bibitem{Concha:2016kdz}
P.~K.~Concha, R.~Durka, C.~Inostroza, N.~Merino and E.~K.~Rodr\'\i{}guez,
``Pure Lovelock gravity and Chern-Simons theory,''
Phys. Rev. D \textbf{94} (2016) no.2, 024055
doi:10.1103/PhysRevD.94.024055
[arXiv:1603.09424 [hep-th]].

\bibitem{Concha:2016tms}
P.~K.~Concha, N.~Merino and E.~K.~Rodr\'\i{}guez,
``Lovelock gravities from Born\textendash{}Infeld gravity theory,''
Phys. Lett. B \textbf{765} (2017), 395-401
doi:10.1016/j.physletb.2016.09.008
[arXiv:1606.07083 [hep-th]].

\bibitem{Concha:2017nca}
P.~Concha and E.~Rodr\'\i{}guez,
``Generalized Pure Lovelock Gravity,''
Phys. Lett. B \textbf{774} (2017), 616-622
doi:10.1016/j.physletb.2017.10.019
[arXiv:1708.08827 [hep-th]].

\bibitem{Fierro:2014lka}
O.~Fierro, F.~Izaurieta, P.~Salgado and O.~Valdivia, \\
``Minimal AdS-Lorentz supergravity in three-dimensions,''
Phys. Lett. B \textbf{788} (2019), 198-205
doi:10.1016/j.physletb.2018.10.066
[arXiv:1401.3697 [hep-th]].

\bibitem{Andrianopoli:2018ymh}
L.~Andrianopoli, B.~L.~Cerchiai, R.~D'Auria and M.~Trigiante,
``Unconventional supersymmetry at the boundary of AdS$_{4}$ supergravity,''
JHEP \textbf{04} (2018), 007
doi:10.1007/JHEP04(2018)007
[arXiv:1801.08081 [hep-th]].

\bibitem{Alvarez:2011gd}
P.~D.~Alvarez, M.~Valenzuela and J.~Zanelli,
``Supersymmetry of a different kind,''
JHEP \textbf{04} (2012), 058
doi:10.1007/JHEP04(2012)058
[arXiv:1109.3944 [hep-th]].

\bibitem{Andrianopoli:2019sqe}
L.~Andrianopoli, B.~L.~Cerchiai, P.~A.~Grassi and M.~Trigiante,
``The Quantum Theory of Chern-Simons Supergravity,''
JHEP \textbf{06} (2019), 036
doi:10.1007/JHEP06(2019)036
[arXiv:1903.04431 [hep-th]].

\bibitem{Hughes:2012vg}
T.~L.~Hughes, R.~G.~Leigh and O.~Parrikar,
``Torsional Anomalies, Hall Viscosity, and Bulk-boundary Correspondence in Topological States,''
Phys.\ Rev.\ D {\bf 88}, no. 2, 025040 (2013),
doi:10.1103/PhysRevD.88.025040
[arXiv:1211.6442 [hep-th]].

\bibitem{Parrikar:2014usa}
O.~Parrikar, T.~L.~Hughes and R.~G.~Leigh,
``Torsion, Parity-odd Response and Anomalies in Topological States,''
Phys.\ Rev.\ D {\bf 90}, no. 10, 105004 (2014),
doi:10.1103/PhysRevD. 90.105004
[arXiv:1407.7043 [cond-mat.mes-hall]].

\bibitem{Ciambelli:2018xat}
L.~Ciambelli, C.~Marteau, A.~C.~Petkou, P.~M.~Petropoulos and K.~Siampos,
``Covariant Galilean versus Carrollian hydrodynamics from relativistic fluids,''
Class. Quant. Grav. \textbf{35} (2018) no.16, 165001
doi:10.1088/1361-6382/aacf1a
[arXiv:1802.05286 [hep-th]].

\bibitem{Ciambelli:2018ojf}
L.~Ciambelli and C.~Marteau,
``Carrollian conservation laws and Ricci-flat gravity,''
Class. Quant. Grav. \textbf{36} (2019) no.8, 085004
doi:10.1088/1361-6382/ab0d37
[arXiv:1810.11037 [hep-th]].

\bibitem{Campoleoni:2018ltl}
A.~Campoleoni, L.~Ciambelli, C.~Marteau, P.~M.~Petropoulos and K.~Siampos,
``Two-dimensional fluids and their holographic duals,''
Nucl. Phys. B \textbf{946} (2019), 114692
doi:10.1016/j.nuclphysb.2019.114692
[arXiv:1812.04019 [hep-th]].

\bibitem{Ciambelli:2018wre}
L.~Ciambelli, C.~Marteau, A.~C.~Petkou,  P.~M.~Petropoulos and K.~Siampos,
``Flat holography and Carrollian fluids,'' \\
JHEP \textbf{07} (2018), 165
doi:10.1007/JHEP07(2018)165
[arXiv:1802.06809 [hep-th]].

\bibitem{Bagchi:2010zz}
A.~Bagchi,
``Correspondence between Asymptotically Flat Spacetimes and \\ Nonrelativistic Conformal Field Theories,'' 
Phys. Rev. Lett. \textbf{105} (2010), 171601
doi:10.1103/PhysRevLett.105.171601
[arXiv:1006.3354 [hep-th]].

\bibitem{Bagchi:2012cy}
A.~Bagchi and R.~Fareghbal,
``BMS/GCA Redux: Towards Flatspace Holography from Non-Relativistic Symmetries,''
JHEP \textbf{10} (2012), 092
doi:10.1007/JHEP10(2012)092
[arXiv:1203.5795 [hep-th]].

\bibitem{Bagchi:2019xfx}
A.~Bagchi, A.~Mehra and P.~Nandi,
``Field Theories with Conformal Carrollian Symmetry,''
JHEP \textbf{05} (2019), 108
doi:10.1007/JHEP05(2019)108
[arXiv:1901.10147 [hep-th]].

\bibitem{Bergshoeff:2015wma}
E.~Bergshoeff, J.~Gomis and L.~Parra,
``The Symmetries of the Carroll Superparticle,''
J. Phys. A \textbf{49} (2016) no.18, 185402
doi:10.1088/1751-8113/49/18/185402
[arXiv:1503.06083 [hep-th]].

\bibitem{Ravera:2019ize}
L.~Ravera,
``AdS Carroll Chern-Simons supergravity in 2 + 1 dimensions and its flat limit,''
Phys. Lett. B \textbf{795} (2019), 331-338
doi:10.1016/j.physletb.2019.06.026
[arXiv:1905.00766 [hep-th]].

\bibitem{Ali:2019jjp}
F.~Ali and L.~Ravera,
``$\mathcal{N}$-extended Chern-Simons Carrollian supergravities in $2+1$ spacetime dimensions,''
JHEP \textbf{02} (2020), 128
doi:10.1007/JHEP02(2020)128
[arXiv:1912.04172 [hep-th]].

\end{thebibliography}
\end{document}